\documentclass[journal]{IEEEtran}

\usepackage{amssymb,amsmath,epsfig,cite,comment,hyperref,verbatim,graphicx,graphics,multirow,bm}
\usepackage{setspace,url}
\usepackage{subfigure}
\usepackage{float}

\usepackage{array, tabularx, boldline}
\usepackage{cellspace}
\title{Analysis and Detection of Pathological Voice using Glottal Source Features}
\author{Sudarsana Reddy Kadiri and Paavo Alku,~\IEEEmembership{Senior Member,~IEEE}
\thanks{S.R. Kadiri and P. Alku are with the Dept. of Signal Processing and Acoustics, Aalto University, Finland, E-mail: \{sudarsana.kadiri;paavo.alku\}@aalto.fi. This study was funded by the Academy of Finland (project no. 284671, 312490).}}

\begin{document}
\maketitle

\begin{abstract}
Automatic detection of voice pathology enables objective assessment and earlier intervention for the diagnosis. This study provides a systematic analysis of glottal source features and investigates their effectiveness in voice pathology detection. % (i.e., normal {\it vs} pathological). 
Glottal source features are extracted using glottal flows estimated with the quasi-closed phase (QCP) glottal inverse filtering method, using approximate glottal source signals computed with the zero frequency filtering (ZFF) method, and using acoustic voice signals directly. In addition, we propose to derive mel-frequency cepstral coefficients (MFCCs) from the glottal source waveforms computed by QCP and ZFF to effectively capture the variations in glottal source spectra of pathological voice.  Experiments were carried out using two databases, the Hospital Universitario Pr{\'i}ncipe de Asturias (HUPA) database and the Saarbr{\"u}cken Voice Disorders (SVD) database. 
Analysis of features revealed that the glottal source contains information that discriminates normal and pathological voice. Pathology detection experiments were carried out using support vector machine (SVM). From the detection experiments it was observed that the performance achieved with the studied glottal source features is comparable or better than that of conventional MFCCs and perceptual linear prediction (PLP) features. The best detection performance was achieved when the glottal source features were combined with the conventional MFCCs and PLP features, which indicates the complementary nature of the features.
\end{abstract}


\section{Introduction}
Speech is produced by exciting a time-varying vocal tract system that consists of various articulators (such as the tongue, jaw, lips) by a time-varying excitation signal. The main purpose of speech in communication is to convey a linguistic message. Apart from linguistic content, speech also contains rich information about the language and dialect as well as about the speaker's gender, age, emotions and state of health. This work studies pathological voice and compares it to normal voice using both analysis and detection (i.e., normal vs. pathological). Voice pathologies arise due to infections, physiological and psychogenic causes and due to vocal misuse that is prevalent in professions such as singers, teachers, and customer service representatives \cite{Disordersbook,williams2003}. Automatic detection of voice pathology enables an objective assessment and an early intervention for the  diagnosis. A typical voice pathology detection system consists of two main stages: the first stage is the representation of the input acoustic speech signal (i.e., feature extraction) and the second stage is the classifier (i.e., normal vs. pathological decision). The main focus of the current study is on the first stage.

Feature sets used for voice pathology detection can be broadly classified into the following three categories \cite{AriasITBE11,GOMEZGARCIA2019}: (1) perturbation measures, (2) spectral and cepstral measures, and (3) complexity measures. Perturbation measures capture the presence of aperiodicity and aspiration noise in the acoustic speech signal that occur due to irregular movement of the vocal folds and incomplete glottal closure. The popular parameters in this category are jitter and shimmer \cite{little2009suitability,Silva2009,jittersty,PerturbationNDA,pertparsa,NFL,MEKYSKArca}. Jitter measures the short-term perturbations of the fundamental frequency ($F_0$) and shimmer measures the short-term perturbations in amplitude \cite{little2009suitability,roy2013evidence}. Several variations of jitter (such as  relative jitter, relative jitter average perturbation and jitter five-point period perturbation quotient) and shimmer (such as absolute shimmer, relative shimmer and shimmer three-point amplitude perturbation quotient) have been used for voice pathology detection \cite{roy2013evidence,GarciaMG19}. The estimation of these features depends on $F_0$, but accurate estimation of $F_0$ is known to be difficult in pathological voice \cite{MANFREDIf0,speccep}. Even though many previous studies have investigated jitter and shimmer features, it is worth observing that these features are not included in the feature set recommended by the American Speech-Language-Hearing Association due to their lack of clinical voice utility \cite{patel2018recommended}. For more details of the recommended acoustic measures, see Table 2 in \cite{patel2018recommended}. Other popular perturbation measures that quantify the presence of aspiration noise include the harmonics-to-noise ratio (HNR) \cite{hnrjasa,hnricassp}, normalized noise entropy (NNE) \cite{kasuya1986normalized}, and glottal-to-noise excitation (GNE) ratio \cite{gne2010,gneparsa,gneacta}. HNR is defined as the ratio between the harmonic component energy and the noise component energy. NNE is the ratio between the energy of noise and the total energy of the signal. GNE measures the correlation between Hilbert envelopes in different frequency bands of the acoustic speech signal.

Measures derived from spectrum and cepstrum have been used extensively for voice pathology detection because these methods are typically easy to compute and do not need the estimation of $F_0$ \cite{specmaveba,cepfeat,speccep}. The popular features in this category are mel-frequency cepstral coefficients (MFCCs) \cite{speccep,mfccann,GOMEZGARCIA2019} that utilize the principles of human auditory processing in the mel-scale and the decorrelating property of cepstrum. In addition, linear predictive cepstral coefficients (LPCCs) \cite{LPfeatpat,featclassifier,gneparsa} and perceptual linear prediction (PLP) \cite{LITTLEnda,GOMEZGARCIA2019} coefficients have been used in voice pathology detection. LPCCs capture the vocal tract system characteristics. PLP features are based on the modelling of the human auditory system using the Bark scale, equal loudness-level curve and intensity-to-loudness conversion \cite{plpextract}. Another popular feature in this category is the cepstral peak prominence (CPP) \cite{hillenbrand1994acoustic,CPPbio}. A larger CPP value indicates a more prominent periodic structure of the signal. A variant of CPP that has been proposed by smoothing the cepstrum and the corresponding parameter is referred to as the smoothed CPP \cite{FraileG14}. Studies \cite{drugman2009mutual,FRAILE201311}
 have additionally used the average spectral energies in low-frequency and high-frequency bands. Features derived from time-frequency decomposition techniques such as adaptive time-frequency transform \cite{tfauma,Ghoraani2009}, wavelet transform \cite{wavelet1,waveletann,fonseca2007wavelet,waveletpacket}, modulation spectrum \cite{MSyan,MarkakiSAG10,modspebio} and empirical mode decomposition \cite{Kaleem2013} have also been investigated for voice pathology detection.

Complexity measures have been proposed to capture properties such as aperiodicity, non-linearity and non-stationarity present in the signal through estimators based on non-linear dynamic analysis \cite{little2007exploiting,little2009suitability,ndareview,henriquez2009characterization,pean2000fractal,tsanaspd1,VAZInda}. It is known that nonlinear phenomena are common in natural physiological systems such as speech production. Non-linear dynamic analysis characterizes the dynamic changes in voice pathologies that occur due to irregular and improper movement of the vocal folds. The popular parameters in this category are computed using the fractal dimension or the correlation dimension \cite{little2007exploiting,TRAVIESnda,giovanni1999determination,tsanaspd,LITTLEnda}. The complex measures investigated in several studies consist of the following: the largest Lyapunov exponent, the recurrence period density entropy, Hurst exponent, detrended fluctuation analysis, approximate entropy, sample entropy, modified sample entropy, Gaussian kernel sample entropy, fuzzy entropy, hidden Markov model (HMM) entropy and Shannon HMM entropy \cite{hmment,GarciaGC12,waveletann,fonseca2007wavelet}. These features capture the dynamic variants/invariants, long-range correlations, regularity or predictability information present in the signal.

It should be noted that the estimation of perturbation features and complexity features depends on the precise estimation of $F_0$ and the selection of the appropriate window length \cite{MANFREDIf0,Jittermeasure}. On the other hand, extraction of spectral or cepstral features does not depend on $F_0$. In \cite{GarciaMG19} and \cite{upm49565}, it was found that voice pathology detection performance with spectral features (such as MFCCs and PLPs) alone is comparable or better than that given by perturbation and complexity features in sustained vowels and continuous speech. More details of the studies on pathological voice and various features used for voice pathology detection can be found in recent review articles \cite{GOMEZGARCIA2019,GarciaMG19}. Regarding classifiers, several known techniques, such as kNN, GMM, LDA, HMM, ANN, CNN and SVM, have been used for pathological voice \cite{mfccgmm,featclassifier,nn,ARJMANDIsvm,chen2007svm,waveletpacket,hmmpat,mfccann,alhussein2018voice,alhussein2019automatic,muhammad2017smart,hossain2016healthcare}. Among the different classifiers, SVM has been found to be the most suitable classifier for voice pathology detection \cite{HEGDE2018}. More details of various classifiers and machine learning techniques used for voice pathology detection can be found in the recent review published in \cite{HEGDE2018}.

Since voice pathologies affect the speech production mechanism, both the glottal source and the vocal tract system need to be represented and parameterized effectively in the analysis and detection of voice pathology. Existing studies have captured the vocal tract system characteristics effectively by deriving spectral or cepstral features such as MFCCs and PLPs. However, there is little previous research on the systematic investigation of glottal source features for the analysis and detection of voice pathologies. In the few studies \cite{glottalpat1,glottalpat2,glottalpat3,MUHAMMAD2017156}, authors have mainly exploited features that capture the specific glottal source characteristics such as HNR, GNE and spectral energies in low-frequency and high-frequency bands of the glottal source.%H1-H2

The current study presents a systematic analysis of glottal source features in normal and pathological voice and investigates their effectiveness in voice pathology detection. The glottal source features are derived from the glottal flow waveforms estimated using the quasi-closed phase (QCP) glottal inverse filtering method \cite{Airaksinen2014} and from the approximate glottal source signals computed by the zero frequency filtering (ZFF) method \cite{Murty32}. The glottal flow signals estimated using QCP are parameterized in terms of time-domain and frequency-domain features \cite{Aparat1,Paavo11}. The features derived from the ZFF method consist of the strength of excitation (SoE), energy of excitation (EoE), loudness measure and ZFF signal energy \cite{Phsing}. In addition to parameterizing glottal source waveforms computed by QCP and ZFF, we also use features which are derived directly from acoustic speech signals and which capture the specific property of the glottal source. These features are the maximum dispersion quotient (MDQ) \cite{MDQ}, peak slope (PS) \cite{PS}, cepstral peak prominence (CPP) \cite{hillenbrand1994acoustic}, and Rd shape parameter \cite{DegottexG2011msp,HuberS2012mspd2ix}. Further, we propose to derive MFCCs from the glottal source waveforms to effectively capture glottal source variations in pathological voice. In total, this results in five sets of glottal source features as follows. 
\begin{itemize}
 \item Time-domain and frequency-domain features derived from the glottal source waveforms estimated with the QCP method
 \item Features derived from the approximate glottal source waveforms computed by the ZFF method
 \item Features which are derived directly from acoustic voice signals and which capture the specific property of the glottal source
 \item MFCCs derived from the glottal flow waveforms estimated with the QCP method
 \item MFCCs derived from the approximate glottal source waveforms given by the ZFF method
\end{itemize}

Voice pathology detection experiments were carried out using two databases, the Hospital Universitario Pr{\'i}ncipe de Asturias (HUPA) database \cite{ModulationSM,MSMFCC} and the Saarbr{\"u}cken Voice Disorders (SVD) database \cite{svddb2,svddb1} that are considered the most reliable and standard databases for voice pathology detection \cite{GOMEZGARCIA2019,GarciaMG19,svduse2}. We did not utilize the popular MEEI database because it suffers from the  problems such as having different recording conditions between healthy and pathological voices (see, e.g., \cite{henriquez2009characterization,gneparsa,AriasITBE11}).  The conventional MFCC and PLP features, which were shown to be effective for voice pathology detection in \cite{GarciaMG19}, are used as the baseline features. Additionally, the complementary nature of the glottal source features is demonstrated, when the glottal source features are combined with the conventional MFCC and PLP features.

The paper is organized as follows. Section~\ref{sec:SPM} describes the two signal processing methods, QCP and ZFF, for deriving glottal source waveforms. The extraction of the glottal source features is discussed in Section~\ref{sec:VSFfeats}.  Section~\ref{sec:analysis} presents the systematic analysis of the glottal source features for normal and pathological voice. Section~\ref{sec:mfccvs} describes the extraction of MFCCs from the glottal source waveforms. Experimental protocol is discussed in Section~\ref{sec:exppro}, which includes the pathology databases, parameters used for feature extraction, baseline features used for comparison, details of the classifier and evaluation metrics. Results and discussion of the detection experiments are presented in Section~\ref{sec:RnD}. Finally, Section~\ref{sec:summary} summarizes the paper.

\section{Signal Processing Methods used for Deriving Glottal Source Waveforms}
\label{sec:SPM}
This section describes the two signal processing methods used in the present study, the QCP glottal inverse filtering method \cite{Airaksinen2014} and the ZFF method \cite{Murty32}, for the estimation of glottal source waveforms. It should be noted that QCP is based on the source-filter model of speech production but ZFF does not use the source-filter model. Hence, these two methods are expected to capture distinct information.

\subsection{The quasi-closed phase (QCP) method}
\label{sec:QCP}
\begin{figure}[htbp]
\centering
\vspace{-0.1cm}
%\hspace{0.8cm}
\includegraphics[width=\columnwidth,height=3.4cm,trim={0cm 6cm 0cm 4.5cm},clip]{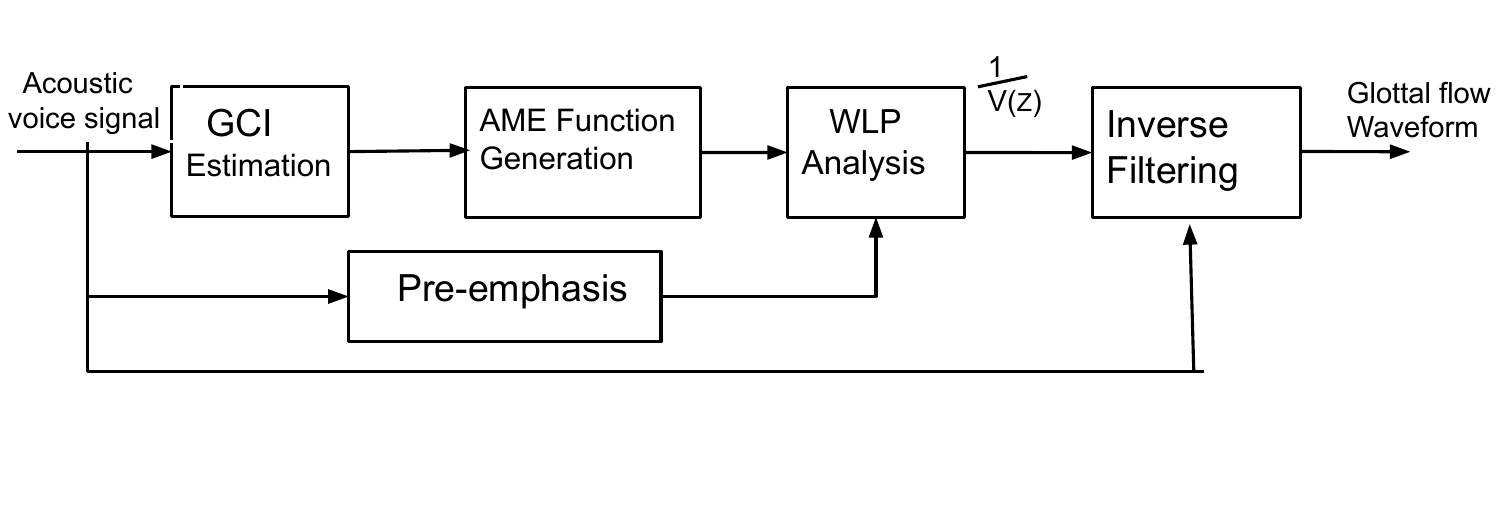}
\vspace{-0.8cm}
\caption{\label{qcpbd}Block diagram of the QCP method.}
\vspace{-0.5cm}
\end{figure}

The QCP method \cite{Airaksinen2014} is a recently proposed glottal inverse filtering method for the automatic estimation of the glottal source waveform from speech. The method is based on the principles of closed phase (CP) \cite{Wong79} analysis which estimates the vocal tract model from few speech samples located in the CP of the glottal cycle using linear prediction (LP) analysis. In contrast to the CP method, QCP takes advantage of all the speech samples of the analysis frame in computing the vocal tract model. This is carried out using weighted linear prediction (WLP) analysis with the attenuated main excitation (AME) \cite{AME} waveform as the weighting function.  The AME function is designed using glottal closure instants (GCIs) and the function attenuates the contribution of the open phase samples in the computation of the acoustic speech signal's covariance or autocorrelation function. This operation results in better estimates of the vocal tract transfer function $V(z)$. Finally, the estimate of the glottal flow waveform is obtained by inverse filtering the input acoustic speech signal with the vocal tract transfer function $V(z)$.  The QCP method was shown to be better than four existing inverse filtering methods in the estimation of the glottal flow from modal and non-modal types of phonation \cite{Airaksinen2014}. This justifies the usage of QCP as a glottal inverse filtering method in the present study. A block diagram describing the steps involved in the QCP method is shown in Fig.~\ref{qcpbd}.

\subsection{The zero frequency filtering (ZFF) method}
The ZFF method was proposed in \cite{Murty32} based on the fact that the effect of an impulse-like excitation (that occurs at the instant of glottal closure) is present throughout the spectrum including the zero frequency, while the vocal tract characteristics are mostly reflected in resonances at much higher frequencies. In this method, the acoustic speech signal is passed through a cascade of two zero frequency resonators and the resulting signal is equivalent to integration of the signal four times. Hence, the output grows or decays as a polynomial function of time. The trend is removed by subtracting the local mean computed over the average pitch period at each sample and the resulting output signal is referred as the zero frequency filtered (ZFF) signal. In this study, we consider the ZFF signal as an approximate glottal source waveform.

The following steps are involved to derive the ZFF signal:
\begin{enumerate}
 \setlength{\itemsep}{-0.1\baselineskip}
\vspace{-0.1cm}
\item The acoustic voice signal ($s[n]$) is first differentiated as follows to remove any low-frequency trend
%\vspace{-0.4cm}
\begin{eqnarray}
x[n]=s[n]-s[n-1]. \label{eqn:preemp}
\end{eqnarray}

\item The differentiated signal is passed through a cascade of two zero frequency resonators (pair of poles on the unit circle along the positive real axis in the $z$-plane). This filtering can be expressed as follows 
%\vspace{-0.4cm}
\begin{eqnarray}
y_o[n]=\sum_{k=1}^{4}{a_k y_o[n-k]} + x[n],
\end{eqnarray}
where $a_1=+4$, $a_2=-6$, $a_3=+4$, $a_4=-1$.
The resulting signal $y_o[n]$ is equivalent to integration (or cumulative sum in the discrete-time domain) of the acoustic voice signal four times, hence it approximately grows or decays as a polynomial function of time. 

\item The trend in $y_o[n]$ is removed by subtracting the local mean computed over the average pitch period (derived using autocorrelation) at each sample. The resulting signal ($y[n]$) is called the ZFF signal and is computed as follows 
%\vspace{-0.3cm}
\begin{eqnarray}
y[n]=y_o[n]-\frac{1}{2N+1}\sum_{i=-N}^{N}{y_o[n+i]}, \label{eqn:trendrem}
\end{eqnarray}
where $2N+1$ corresponds to the number of samples in the window used for trend removal. 
\end{enumerate}

The ZFF signal is used to derive the glottal source characteristics \cite{Murty32}.  The positive-to-negative zero-crossings (PNZCs) correspond to GCIs (or epochs) by considering the negative polarity of the signal \cite{Murty32,myspd}. Let us denote epochs by ${\cal{E}}=\{e_1,e_2,...,e_M\}$, where $M$ is the number of epochs. The time duration between any two adjacent epochs gives the instantaneous fundamental period ($T_0[k]$), and its reciprocal gives the instantaneous fundamental frequency ($F_0[k]$), i.e.,
\begin{eqnarray}
T_0[k]&=&\frac{(e_k-e_{k-1})}{f_s}, \qquad k=2,3,...,M, \\
F_0[k]&=&\frac{1}{T_0[k]} = \frac{f_s}{(e_k-e_{k-1})}, \qquad k=2,3,...,M,
\end{eqnarray}
where $f_s$ is the sampling frequency.

\begin{figure}[h]
\centering
\begin{center}
 \vspace{-0.2cm}
%  \hspace{0.8cm}
\includegraphics[width=\columnwidth,height=3.3cm,trim={0cm 0cm 0cm 0cm},clip]{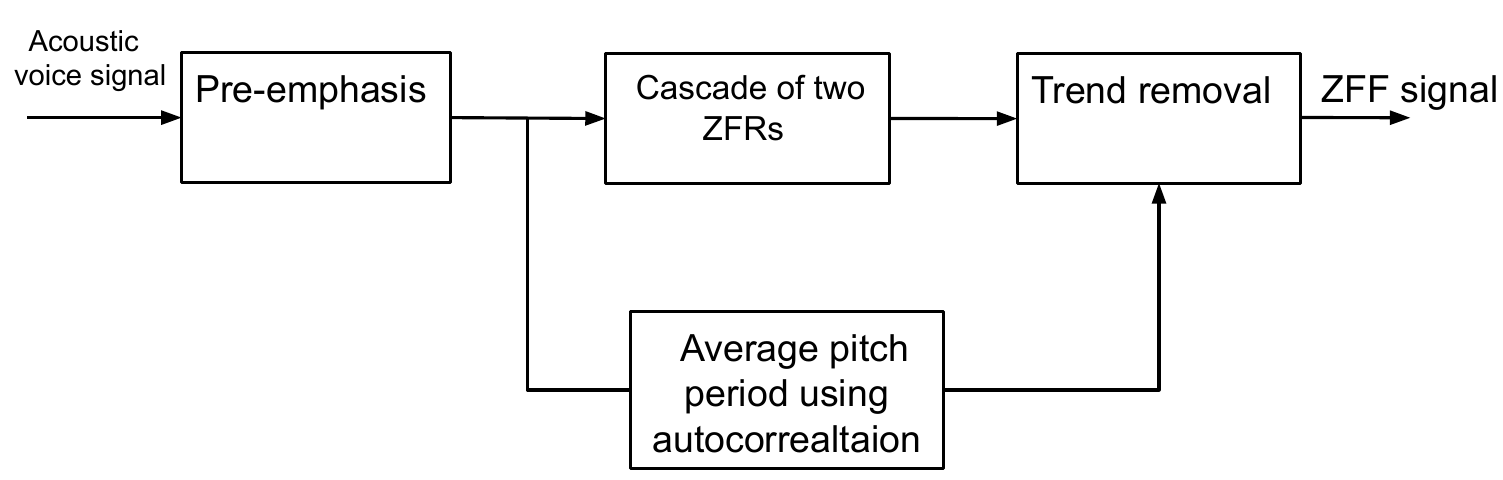}
\vspace{-.3cm}
 \caption{\label{zffbd}Block diagram of the ZFF method.}
 \vspace{-.1cm}
 \end{center}
\end{figure}

Another interesting property of the ZFF signal is that the slope of the signal around each PNZC is proportional to the rate of closure of the vocal folds as measured using differentiated electroglottography (EGG) signals at the instants of glottal closure. A block diagram describing the steps involved in the ZFF method is shown in Fig.~\ref{zffbd}.  

\begin{figure}[h]
%\centering
\begin{center}
\includegraphics[width=\columnwidth, height=6.7cm,trim={0cm 0cm 0.3cm 0.2cm},clip]{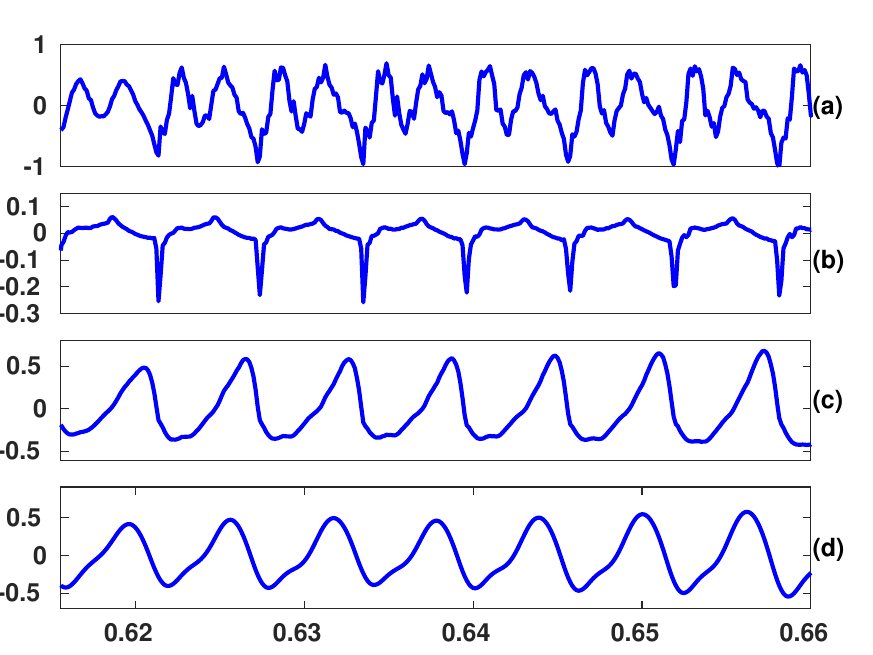}%{zff_gf_gfd1.eps}%
\begin{minipage}{0.5cm}
\vspace{0.7cm}
\hspace*{-5cm}{{\bf Time (sec)}}
\end{minipage}
\caption{Illustration of glottal source waveforms derived using the QCP and ZFF methods: (a) acoustic speech signal, (b) differentiated EGG signal, (c) glottal source waveform estimated by QCP, and (d) approximate glottal source waveform estimated by ZFF (reversed in polarity for visualization purpose).} 
\label{zff_feats}
\end{center}
\end{figure}

To illustrate the glottal source waveforms computed by QCP and ZFF, a segment of voiced speech along with the simultaneously recorded EGG signal from the CMU ARCTIC database \cite{arcticDatabase} is used. Fig.~\ref{zff_feats}(a) and Fig.~\ref{zff_feats}(b) show the acoustic speech signal and the differentiated EGG, respectively. Glottal source waveforms computed by QCP and ZFF are shown in Fig.~\ref{zff_feats}(c) and Fig.~\ref{zff_feats}(d), respectively. 

\section{Extraction of Glottal Source Features}
\label{sec:VSFfeats}
This section describes the extraction of features from the glottal source waveforms computed using QCP and ZFF. In addition, the section explains the extraction of the glottal source features that are computed directly from the acoustic voice signal and that capture specific properties of the glottal source.

\subsection{Glottal source features derived using the QCP method} 
In order to represent the glottal flow waveform in a compact form, different methods have been developed for parameterization and they can be grouped into two categories: time-domain and frequency-domain glottal features (also called glottal parameters).

\subsubsection{Time-domain glottal features} 
Time-domain glottal flow signals can be parameterized using time-based and amplitude-based features \cite{Aparat1,Paavo11}. In the case of time-based features, the most classical approach is to compute time-duration ratios between the different phases (closed phase, opening phase, and closing phase) of the glottal source waveform in a glottal cycle. The measures are defined by extracting critical time instants (such as the instant of glottal closure, primary and secondary glottal opening, the instant of minimum and maximum glottal flow) from the glottal source waveform. In the case of amplitude-based features (amplitude quotient  \cite{Alku96,Alku2006amplitude} and normalized amplitude quotient \cite{Alku02}), the amplitude of the glottal flow and its derivative are used \cite{Fant95, Alku96, Alku02}. The normalized amplitude quotient has been shown to be a strong correlate of the closing quotient, and it has been extensively used in analyzing voice quality \cite{Alku02}. Extraction of critical time instants is often difficult and to overcome this, sometimes time-based features are computed by replacing the true closure and opening instants by the time instants when the glottal flow crosses a level, which is set to a value between the maximum and minimum amplitude of glottal flow in a glottal cycle \cite{Paavo11}.

\subsubsection{Frequency-domain glottal features} 
While the computation of time-domain features from the glottal source waveform is straightforward, these features are affected by distortions such as formant ripple due to incomplete canceling of formants by the inverse filter \cite{Paavo11}. In such cases, it is useful to derive frequency-domain features for the glottal source waveform. Frequency-domain features are computed from the spectrum of the glottal source and they essentially measure the slope of the spectrum. Several studies have quantified the spectral slope of the glottal source by utilizing the level of  $F_0$ and its harmonics. The most widely used features are the amplitude difference between $F_0$ and the first harmonic (H1-H2) \cite{Titze92}, the harmonic richness factor (HRF) \cite{VQ_ASP}, and the parabolic spectral parameter (PSP) \cite{PSP}. HRF is the ratio between the sum of the amplitudes of the harmonics above  $F_0$ and the amplitude of $F_0$. PSP is derived by fitting a parabola to the low frequencies of the glottal flow spectrum \cite{PSP}. 

A total of 12 glottal features (9 time-domain and 3 frequency-domain features) defined in \cite{Aparat1} are used in this study to characterize the glottal flow waveforms estimated by the QCP glottal inverse filtering method. These features are extracted using the APARAT Toolbox \cite{Aparat1} and they are listed in Table~\ref{TD_FD}.

\begin{table}[h!]
\caption{{Time-domain and frequency-domain glottal features derived from glottal flows estimated by QCP.}}
\vspace{-1mm}
\centering
\label{TD_FD} 
%\resizebox{8.4cm}{2.8cm}{
\small\addtolength{\tabcolsep}{-6pt}
\centering
\begin{tabularx}{8.8cm}{|SlV{2}X|} \hlineB{1.2} % {|l|c|} \hline
  & {\bf {~~~~Time-domain features}}\\ \hline  \hline 
OQ1   		&~Open quotient, calculated from the primary glottal opening  \\  \hline 
OQ2   		&~Open quotient, calculated from the secondary glottal opening  \\  \hline 
NAQ   		&~Normalized amplitude quotient  \\  \hline 
AQ        	&~Amplitude quotient \\  \hline 
ClQ 		&~Closing quotient \\   \hline 
OQa 		&~Open quotient, derived from the LF model \\  \hline 
QoQ         &~Quasi-open quotient \\  \hline 
SQ1  		&~Speed quotient, calculated from the primary glottal opening \\  \hline 
SQ2  		&~Speed quotient, calculated from the secondary glottal opening \\ \hline  \hline 
	        & {\bf~~~~Frequency-domain features} \\ \hline 
H1-H2 		&~Amplitude difference between the first two glottal harmonics \\  \hline 
PSP         &~Parabolic spectral parameter \\  \hline 
HRF  		&~Harmonic richness factor \\	 \hline            
\end{tabularx}
\end{table}

\subsection{Glottal source features derived using the ZFF method} 
From the ZFF method, the following glottal source features are extracted: the strength of excitation (SoE), energy of excitation (EoE), loudness measure and ZFF signal energy. These features have been shown to be useful for discriminating phonation types and emotions in \cite{Phsing,Emo8}. 
 
The four ZFF-based parameters are computed as follows.

\subsubsection{Strength of excitation (SoE)}
The slope of the ZFF signal around each PNZC corresponds to the SoE, which is proportional to the rate of closure of the vocal folds \cite{Emo8}. A measure of SoE around the GCI is given by
\begin{equation}
 SoE=|y[e_k+1]-y[e_k-1]|, \qquad k=1,2,...,M.
\end{equation}
where $y[n]$ is the ZFF signal (Eq.~\ref{eqn:trendrem}). % after trend removal operation (Eq.~\ref{eqn:trendrem}).
\subsubsection{Energy of excitation ($EoE$)}
The $EoE$ feature is computed from the samples of the Hilbert envelope ($h_e[i]$) of the LP residual over a 1-ms region around each GCI. This feature, defined below in Eq.~\ref{eqn:EoE}, has been shown to measure vocal effort \cite{Emo8}.
\begin{eqnarray}
EoE=\frac{1}{2K+1}\sum_{i=-K}^{K}{h_e^2[i]}, \label{eqn:EoE}
\end{eqnarray}
where 2K+1 corresponds to the number of samples in the 1-ms window. 
\subsubsection{Loudness measure}
The loudness measure captures the abruptness of glottal closure \cite{Emo8}, and it is defined according to Eq.~\ref{eqn:loudness} as the ratio between the standard deviation ($\sigma$) and mean ($\mu$) of the samples of the LP residual's Hilbert envelope in a 1-ms region around GCI. %A measure of loudness around the GCI is given by
\begin{eqnarray}
Loudness~measure=\frac{\sigma} {\mu}. \label{eqn:loudness}
\end{eqnarray}
\subsubsection{ZFF signal energy ($v_{zff}[n]$)}
The energy of the ZFF signal is given by
\begin{eqnarray}
v_{zff}[n]=\frac{1}{L}\sum_{i=-L/2}^{L/2}{y^2[n+i]},
\end{eqnarray}
where $y[n]$ is the ZFF signal. The energy of the ZFF signal at GCI is used in this study.
% , and $L$ is the window length (10 ms) over which the energy is computed

\begin{figure}[h]
    \begin{center}
        \begin{tabular}{c}
            {\resizebox*{8.6cm}{5.4cm}{\includegraphics{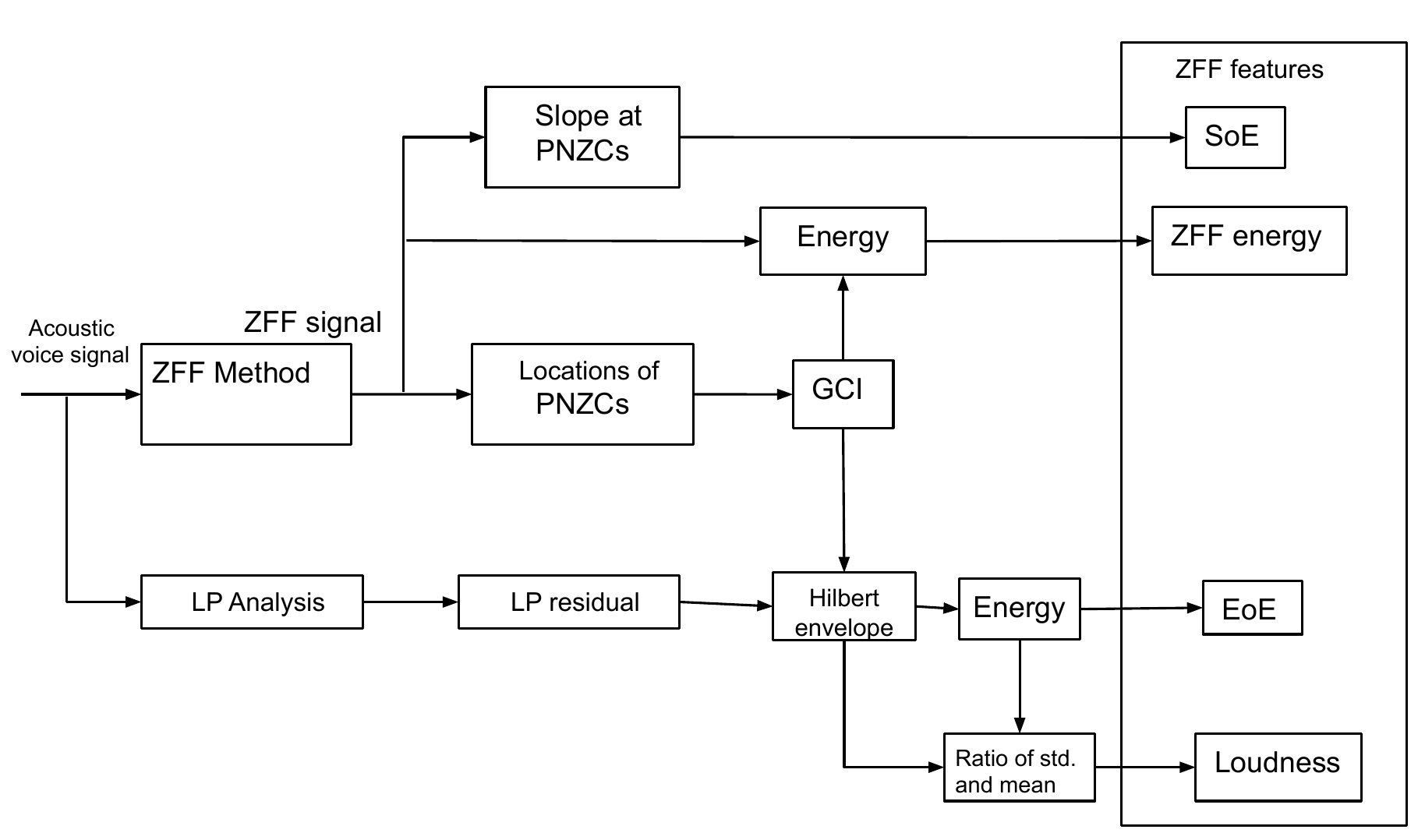}}}
        \end{tabular}
        \renewcommand{\baselinestretch}{1.0}{\parbox{8cm}{
        \caption{\label{fig:zfffeat}{Schematic block diagram for the extraction of glottal source features using the ZFF method.}}}}
    \end{center}
    \vspace{-0.6cm}
\end{figure}

The steps involved in the extraction of glottal source features from the ZFF method are shown in the schematic block diagram in Fig.~\ref{fig:zfffeat}.
%%%%%%%%%%%%%%%%%%%%%%%%%%%%%%%%%%%%%%%%%%%%%%%%%%%%%%%%%%%%%%%%%%%%%%%%%%%%%%%
%%%%%%%%%%%%%%%%%%%%%%%%%%%%%%%%%%%%%%%%%%%%%%%%%%%%%%%%%%%%%%%%%%%%%%%%%%%%%%%

\subsection{Glottal source features derived directly from acoustic voice signals} 
The following four parameters, which capture the specific property of the glottal source, are computed directly from acoustic voice signals without computing the glottal source waveform.
\subsubsection{Maximum dispersion quotient (MDQ)}
The MDQ parameter captures the abruptness of closure of the vocal folds \cite{MDQ}. This parameter measures the dispersion in the LP residual around GCI. Here, wavelet decomposition is carried out for the LP residual. Within a search interval near the GCI of the decomposed signals, the distance of the maxima locations to the given GCI is measured. The average of these distances normalized to the pitch period is referred to as MDQ.

\subsubsection{Peak slope (PS)}
The PS parameter \cite{PS} captures the spectral slope of the glottal source. The method involves computing a wavelet-based decomposition of the acoustic voice signal into octave bands and then fitting a regression line to the maximum amplitudes at the different frequency bands. The slope coefficient of the fitted regression line is referred to as PS. 

\subsubsection{Cepstral peak prominence (CPP)}
The CPP parameter measures the amount of periodicity present in the signal using cepstrum \cite{hillenbrand1994acoustic}. A high CPP value reflects a more periodic structure in a signal. 
Initially this parameter was proposed to characterize the breathiness of voice signals \cite{hillerbrand_breathy}. CPP measures the difference between the most prominent cepstral peak (first rahmonic) and the point with the same quefrency on the regression line through the smoothed cepstrum.

\subsubsection{Rd shape parameter}
The Rd shape parameter is based on first presenting the entire glottal flow waveform using the parametric Liljencrants-Fant (LF) model \cite{Fant95}) and then presenting the LF pulse using a single parameter.  The Rd shape parameter \cite{DegottexG2011msp,HuberS2012mspd2ix} provides a single feature which captures most of the covariation of the LF parameters. A high value of Rd indicates a more relaxed voice.

\section{Analysis of Normal and Pathological Voice with Glottal Source Features}
\label{sec:analysis}
This section presents results that were obtained when the glottal source features described in Section~\ref{sec:VSFfeats} were used to analyze normal and pathological voice. The analyses were carried out using the twenty speakers of HUPA database (details of the database are given in Section~\ref{sec:exppro_db_hupa}). The results obtained are described in feature distributions that are depicted using box plots. 
By presenting box plots of the feature distributions, our aim is to analyze potential differences between different glottal source features in their discriminability of normal and pathological voice. 

\subsection{Analysis of the glottal source features derived using the QCP method}
\label{sec:qcpanalysis}
\begin{figure}[h]
\centering
\includegraphics[width=8.4cm, height=9cm,trim= 1.7cm 1cm 2cm 1cm]{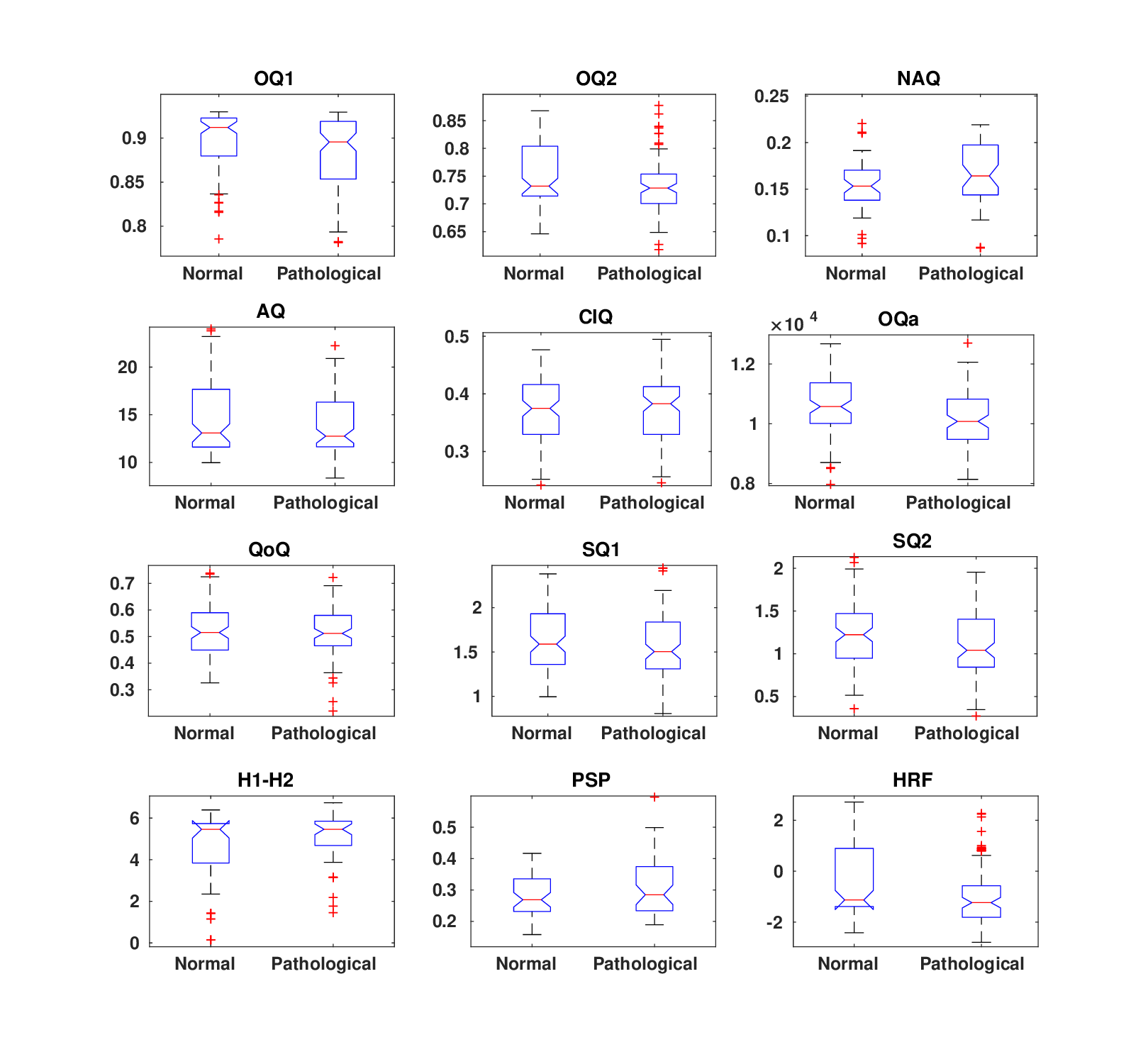}
\caption{\label{qcpanal}Distribution of the glottal source features derived from the QCP method for normal and pathological voice using box plots. The central mark indicates the median, and the bottom and top edges of the box indicate the $25^{th}$ and $75^{th}$ percentiles, respectively. The whiskers on either side cover all points within 1.5 times the interquartile range, and points beyond these whiskers are plotted as outliers using the $'+'$ symbol.}  
\vspace{-0.2cm}
\end{figure}

Figure~\ref{qcpanal} shows distributions of the glottal source features derived using the QCP method for normal and pathological voice. The figure shows the nine time-domain features (rows 1, 2 and 3) and the three frequency-domain features (row 4). It can be seen that the frequency-domain features result in better discrimination of normal and pathological voice compared to the time-domain features. In the time-domain features, NAQ discriminates normal and pathological voice better than the other features. For the open quotients, OQ1 and OQ2 show larger variations in pathological voice compared to normal speech, and QoQ indicates less discriminability. On the other hand, the LF model-based open quotient (OQa) shows good discriminability. 
AQ, ClQ, SQ1 and SQ2 show in general small differences in distributions between normal and pathological voice. This may be due to the difficulty in identifying critical glottal time instants (instant of glottal closure, primary and secondary glottal opening).

The frequency-domain features show in general better discriminability of normal and pathological voice compared to the time-domain features. The values of H1-H2 and PSP are higher in pathological voice compared to normal indicating that the spectral slope of the glottal source is deeper. On the other hand, HRF values are lower for pathological voice due to a weaker harmonic structure (see Fig.~\ref{spectrograms}) in their glottal source spectrum. 

\subsection{Analysis of the glottal source features derived using the ZFF method}
\begin{figure}[h]
%\vspace{-0.6cm}
\centering
\includegraphics[width=7cm, height=6cm,trim= 1.5cm 0.4cm 1.5cm 0.5cm]{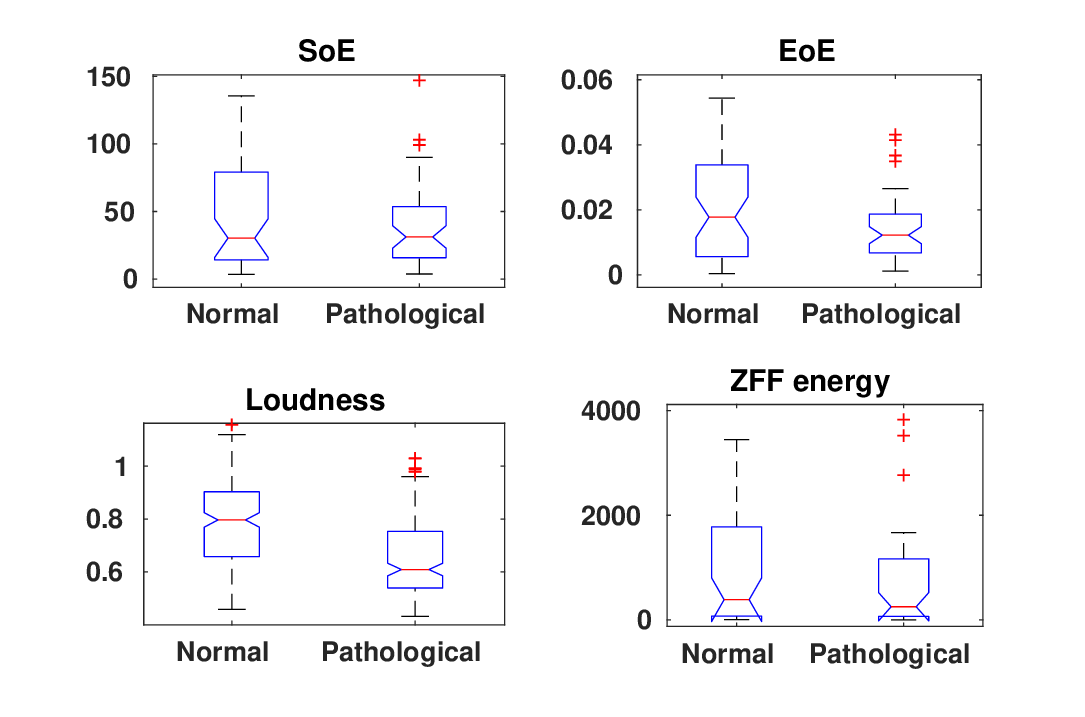}
\caption{\label{zffanal}Distribution of the glottal source features derived from the ZFF method for normal and pathological voice.The central mark indicates the median, and the bottom and top edges of the box indicate the $25^{th}$ and $75^{th}$ percentiles, respectively. The whiskers on either side cover all points within 1.5 times the interquartile range, and points beyond these whiskers are plotted as outliers using the $'+'$ symbol.}  
\vspace{-0.6cm}
\end{figure}

Figure~\ref{zffanal} shows the distribution of the glottal source features derived using the ZFF method. It can be seen that all the features show in general good discriminability of normal and pathological voice. SoE, which measures the strength of the impulse-like excitation at glottal closure, is lower in pathology indicating less abrupt closure of the vocal folds compared to normal speech. EoE, which measures the energy of excitation at the glottal closure and captures the vocal effort required to produce the voice signal, is also lower in pathology compared to normal. %The loudness feature captures the abrupt closure of the vocal folds at the glottis. 
As pathological voice is produced with improper and slower glottal closure, the loudness measure values are also lower. The ZFF signal energy of pathological voice is lower than in normal voice, similar to EoE.  

\subsection{Analysis of the glottal source features derived from acoustic voice signals}
\begin{figure}[h]
%\vspace{-0.6cm}
\centering
\includegraphics[width=7cm, height=6cm,trim= 1.5cm 0.4cm 1.5cm 0.8cm]{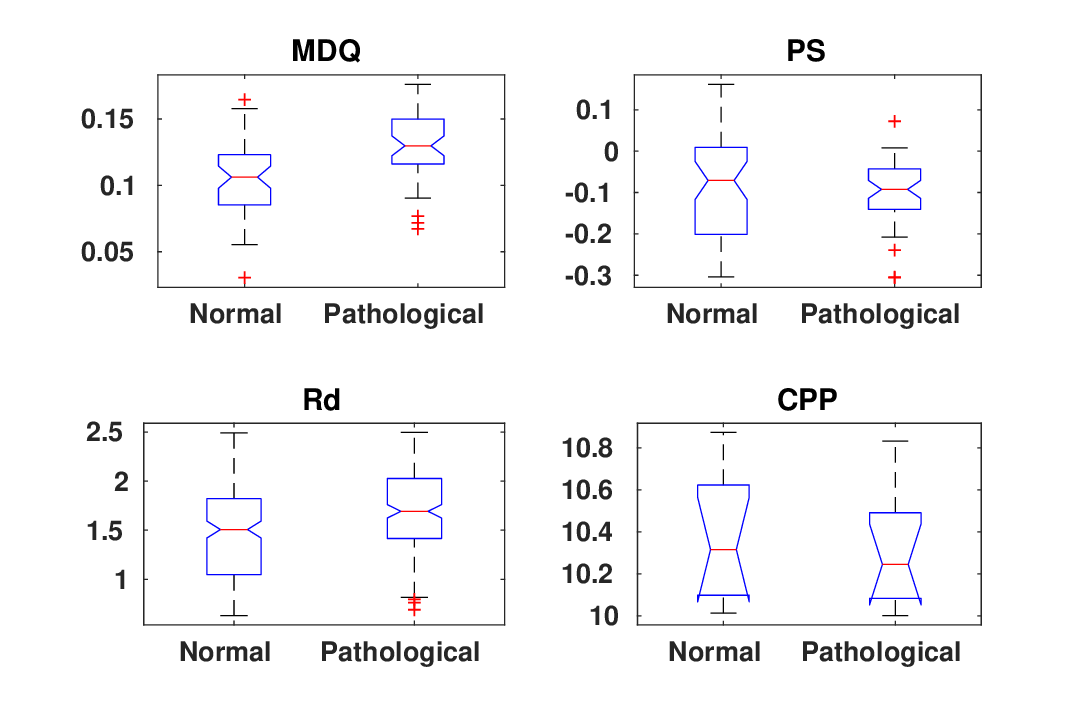}
\caption{\label{othanal}Distribution of the glottal source features derived directly from acoustic speech signals for normal and pathological voice. The central mark indicates the median, and the bottom and top edges of the box indicate the $25^{th}$ and $75^{th}$ percentiles, respectively. The whiskers on either side cover all points within 1.5 times the interquartile range, and points beyond these whiskers are plotted as outliers using the $'+'$ symbol.}  
\vspace{-0.45cm}
\end{figure}

Figure~\ref{othanal} shows the distribution of the glottal source features derived directly from acoustic voice signals. It can be seen that all the features are capable of discriminating normal and pathological voice. The MDQ feature, which measures the dispersion of the LP residual around glottal closure, is high in pathology indicating the occurrence of improper glottal closure and increased aspiration noise. 
PS, which captures the spectral slope  of the glottal source, is also higher in pathology. This observation is not very evident from the figure, as the range of the PS values is higher in normal voice. %The Rd feature captures shape of the glottal flow waveform. 
As pathological voice is produced with a larger amount of aspiration noise and improper glottal closure, it is similar to normal voice of breathy phonation. For breathy or relaxed voices, the Rd feature values are high \cite{DegottexG2011msp,HuberS2012mspd2ix}.
This observation is evident from the box plot of the Rd feature. The CPP feature measures the amount of periodicity present in the signal. Because of improper glottal closure and an increased amount of aspiration noise in pathological voice, the harmonicity of the glottal source spectrum is weaker (see Fig.~\ref{spectrograms}). Hence, the CPP values are lower for pathological voice, which is evident from the box plot.   

\section{Extraction of MFCCs from Glottal Source Waveforms}
\label{sec:mfccvs}
\begin{figure}[h]
%\vspace{-0.2cm}
%\hspace{0.8cm}
\includegraphics[width=8.8cm,height=4.4cm,trim={0cm 0cm 1.6cm 0cm},clip]{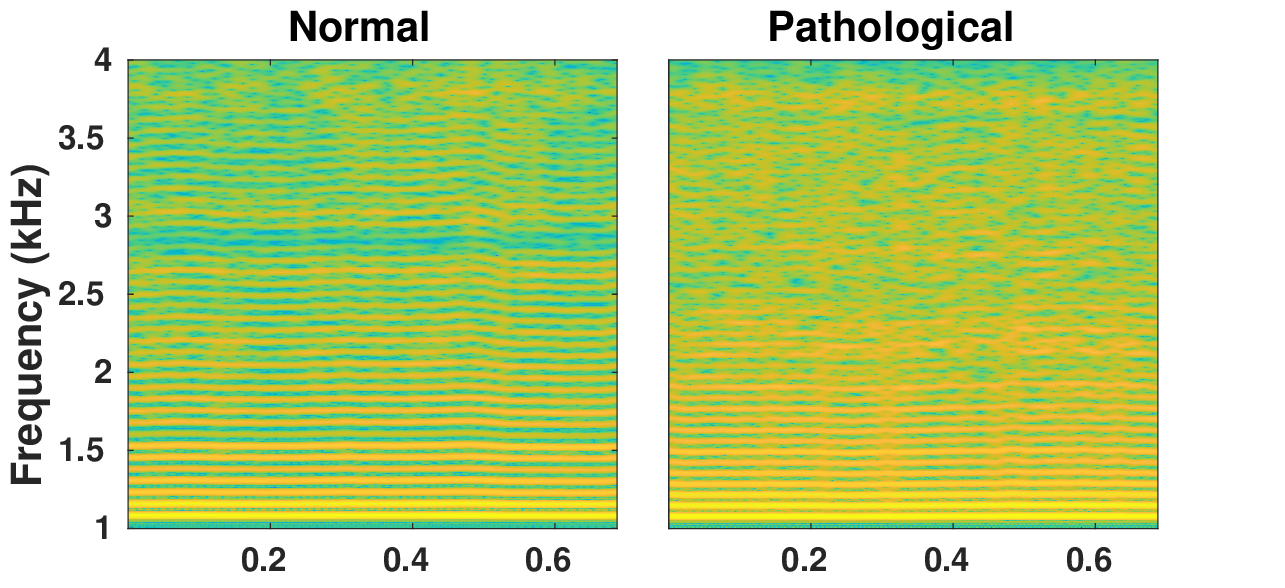}
\begin{minipage}{0.1cm}
\vspace{-0.1cm}
\hspace*{4cm}{{\bf Time~(sec)}}
\end{minipage}
\caption{\label{spectrograms}Illustration of spectrograms of glottal source waveforms estimated using the QCP method for normal and pathological voice.}
%\vspace{-.3cm}
\end{figure}

From the analysis of the glottal source features described in Section~\ref{sec:qcpanalysis}, it can be concluded that the features derived from the glottal source spectrum (frequency-domain features) have a better discriminability compared to the time-domain features. This motivates us to use the entire glottal source spectrum, instead of a few single features, in voice pathology detection. Figure~\ref{spectrograms} shows spectrograms of glottal flow waveforms for normal and pathological voice estimated using the QCP method. It can be seen that there are large variations especially in the harmonic structure of the glottal flow spectra between normal and pathological voice. In order to capture these variations and to represent them in a compact form, we propose to derive MFCCs from the spectra of the glottal source waveforms (as in our recent conference paper \cite{Kadiri2019}). It should be noted that the proposed MFCC feature extraction is similar to the computation of conventional MFCC features, except that the proposed approach operates on the glottal source waveform instead of the acoustic voice signal.  

\begin{figure*}[htbp]
\vspace{-0.1cm}
%\hspace{0.8cm}
\includegraphics[width=18cm,height=2.1cm,trim={0cm 0cm 0cm 0cm},clip]{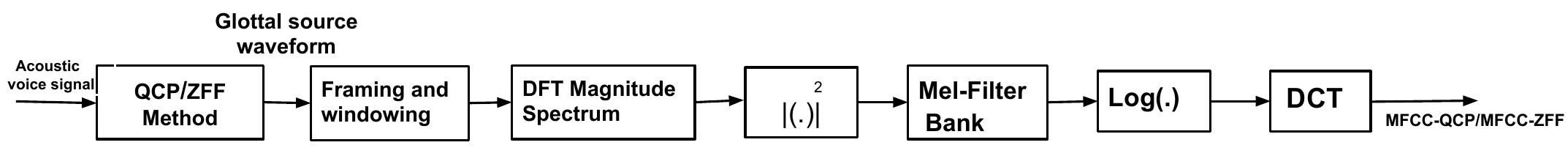}
\vspace{-.1cm}
\caption{\label{mfcc}Extraction of MFCCs from the glottal source waveforms computed by the QCP and ZFF methods.}
\vspace{-.3cm}
\end{figure*}

A schematic block diagram of the extraction of MFCCs from the glottal source waveform given by QCP and ZFF is shown in Fig.~\ref{mfcc}. The method involves short-term spectral analysis, where the glottal source waveform is split into overlapping time-frames and the spectrum of each frame is computed with DFT. The spectrum is estimated using a 1024-point DFT with Hamming windowing in 25-ms frames with a 5-ms shift. Mel-cepstrum is derived from the mel-scale-based analysis of the spectrum of the glottal source, followed by logarithm and discrete cosine transform (DCT). From the entire mel-cepstrum, the first 13 coefficients (including the $0^{th}$ coefficient) are considered for each frame. The resulting cepstral coefficients are referred as MFCC-QCP and MFCC-ZFF for the glottal source waveforms computed by QCP and ZFF, respectively. From static cepstral coefficients, delta and double-delta coefficients are also computed.

\section{Experimental Protocol}
\label{sec:exppro}
This section describes the databases, the feature sets used in voice pathology detection including the baseline features, the classifier and the evaluation metrics.

\subsection{Databases of pathological voice}
\label{sec:exppro_db}
Two databases containing normal and pathological voice are used in this study. These databases are the Hospital Universitario Pr{\'i}ncipe de Asturias (HUPA) database \cite{ModulationSM,MSMFCC} and the Saarbr{\"u}cken Voice Disorders (SVD) database \cite{svddb2,svddb1}.

\subsubsection{The HUPA database}
\label{sec:exppro_db_hupa}
This database was recorded at the Pr{\'i}ncipe de Asturias hospital in Alcal{\'a} de Henares, Madrid, Spain \cite{ModulationSM,MSMFCC}. The dataset contains sustained phonations of the vowel /a/ by 439 adult Spanish speakers (239 healthy and 200 pathological). Originally, the data was recorded with a sampling frequency of 50 kHz and later downsampled to 25 kHz. Pathological voices contain a wide variety of organic pathologies such as nodules, polyps, oedemas and carcinomas. More details of the database can be found in \cite{ModulationSM,MSMFCC,WPCVox}.

\subsubsection{The SVD database}
This database was recorded at the Institut f{\"u}r Phonetik at Saarland University and the Phoniatry Section of the Caritas Clinic St. Theresia in Saarbr{\"u}cken, Germany \cite{svddb2,svddb1}. The data comprises recordings of sustained phonations of the vowels /a/, /i/ and /u/ in normal, high and low pitches, as well as with rising-falling pitch. In addition, the data contains recordings of the sentence "Guten Morgen, wie geht es Ihnen?" ("Good morning, how are you?"). The dataset was recorded from 2225 German speakers, of which 869 are healthy and 1356 are pathological. The database contains 71 different pathologies including both functional and organic pathologies. The data was recorded with a sampling frequency of 50 kHz. As in \cite{GarciaMG19}, in this study we use the vowels /a/, /i/ and /u/, produced using normal pitch, and the running speech after removing the samples with a lower dynamic range, samples which are recorded after voice therapy  and  surgical  intervention. This procedure resulted in data of 1518 speakers, of which 661 are healthy and 857 are pathological. More details of the database can be found in \cite{svddb2,svddb1}.

\subsection{The proposed glottal source feature sets and parameters used for feature extraction}
\label{proposed_feats}
In total, five sets of glottal source features are investigated as listed below: 
\begin{itemize}
 \item Time-domain (OQ1, OQ2, NAQ, ClQ, SQ1, SQ2, AQ, QoQ, OQa) and frequency-domain (H1-H2, PSP, HRF) features derived from the glottal source waveforms computed by the QCP method. These features are extracted for every glottal cycle and QCP analysis is carried out in Hamming-windowed 25-ms frames with a 5-ms shift.   
 
 \item Features derived from the approximate glottal source signals computed by the ZFF method (SoE, EoE, loudness measure, ZFF energy). All these features are computed around GCIs. EoE and loudness measure are computed from the samples of the Hilbert envelope of the LP residual (computed with $12^{th}$ order) over a 1-ms region around each GCI. %SoE and ZFF energy are derived at GCI. 
 
 \item Features that capture the specific property of the glottal source computed directly from acoustic voice signals without computing the glottal source waveform (MDQ, PS, CPP, Rd). All these features are computed in 25-ms Hamming-windowed frames with a 5-ms shift. 
 
 \item MFCC-QCP is computed from the glottal flow waveforms estimated by QCP in 25-ms Hamming windowed frames with a 5-ms shift. First 13 static cepstral coefficients and their delta \& double-delta coefficients are computed yielding a 39-dimensional feature vector.
 
 \item MFCC-ZFF is computed from the approximate glottal source waveforms given by ZFF using 25-ms Hamming windowed frames with a 5-ms shift. Here also, static coefficients and their delta \& double-delta coefficients are computed yielding a 39-dimensional feature vector.
\end{itemize}

\subsection{Baseline features used for comparison}
\label{baseline_feats}
We consider conventional MFCC and PLP features for comparison as they were shown in \cite{GarciaMG19} to provide good discrimination between normal and pathological voice.

\subsubsection{Mel-frequency cepstral coefficients (MFCCs)}
Conventional MFCC features were computed using 25-ms Hamming-windowed frames with a 5-ms shift. The first 13 cepstral coefficients (including the $0^{th}$ coefficient) and their delta \& double-delta coefficients were computed yielding a 39-dimensional feature vector.

\subsubsection{Perceptual linear prediction (PLP) coefficients}
Conventional PLP features were computed using 25-ms Hamming-windowed frames with a 5-ms shift. The first 13 cepstral coefficients (including the $0^{th}$ coefficient) and their delta \& double-delta coefficients were computed yielding a 39-dimensional feature vector.

\subsection{Classifier}
The most popular classifier for voice pathology detection is support vector machine (SVM). In the current study, we use SVM with a radial basis function (RBF) kernel. Experiments were conducted with 20-fold cross-validation, where the data was partitioned randomly into 20 equal portions. One fold was held out to be used for testing with the remaining nineteen folds for training. The training data were z-score normalized and the testing data were normalized by subtracting the mean and dividing by the standard deviation of the training sets for each feature. Evaluation metrics were saved in each fold and this process was repeated for each of the 20 folds. Finally, the evaluation metrics were averaged over the 20 folds for evaluation.

\subsection{Evaluation metrics}
Standard performance metrics for a binary classifier are considered for each one of the aforementioned feature sets \cite{GarciaMG19,ROC}. Therefore, the following metrics are used: accuracy (ACC), sensitivity (SE), specificity (SP), area under the receiver operating characteristic curve (AUC), and equal error rate (EER). For a better performing system, the values of first four metrics should be higher and the last metric should be lower.

\section{Pathology Detection Experiments}
\label{sec:RnD}
Pathology detection experiments were carried out using the SVM classifier with the individual feature sets described in Sections~\ref{proposed_feats}~and~\ref{baseline_feats} as well as with combinations of the feature sets to analyze the complementary information among the features. In the combination of features, the complementary information among the glottal source feature sets and complementary information with the existing spectral features was also investigated. In total, 12 feature sets were investigated, out of which seven were individual feature sets (denoted FS-1 to FS-7) and five were combination of feature sets (denoted FS-8 to FS-12). These feature sets are listed below.

\begin{itemize}
 \item FS-1: OQ1, OQ2, NAQ, ClQ, SQ1, SQ2, AQ, QoQ, OQa, H1-H2, PSP and HRF.
 \item FS-2: SoE, EoE, Loudness, ZFF energy.
 \item FS-3: MDQ, PS, CPP and Rd. 
 \item FS-4: MFCC-QCP% = MFCCs derived from glottal source waveform using QCP method. 
 \item FS-5: MFCC-ZFF% = MFCCs derived from glottal source waveform using ZFF method. 
 \item FS-6: Conventional MFCC features.% = Conventional MFCCs derived from acoustic speech signal. 
 \item FS-7: Conventional PLP features. 
 \item FS-8: Combination of FS-1, FS-2 and FS-3
 \item FS-9: Combination of FS-4 and FS-5.
 \item FS-10: Combination of FS-1, FS-2, FS-3, FS-4 and FS-5 (combination of all glottal source features).
 \item FS-11: Combination of FS-6 and FS-7 (combination of spectral features i.e., MFCC and PLP).
 \item FS-12: Combination of FS-1, FS-2, FS-3, FS-4, FS-5, FS-6, and FS-7 (combination of all glottal source features and spectral features).
\end{itemize}

A total of five experiments were carried out: one experiment using the HUPA dataset (sustained phonation of the vowel /a/) and four experiments using the SVD dataset (sustained phonation of the vowels /a/, /i/, /u/, and the sentence sample), and the corresponding results are given in Tables \ref{HUPA_a} to \ref{SVD_rs}.

Table~\ref{HUPA_a} shows the voice pathology detection results computed for the vowel /a/ in the HUPA database with the individual feature sets (FS-1 to FS-7) and combination of feature sets (FS-8 to FS-12). From the table, it can be observed that in the case of individual feature sets, the feature set FS-5 (MFCC-ZFF) provided the best performance in terms of accuracy (72.89\%), AUC (0.78) and EER (0.253). In terms of AUC and EER, the next best feature set was FS-4 (MFCC-QCP), which provided an AUC of 0.77 and EER of 0.267. The MFCC and PLP feature sets (FS-6 and FS-7) were also close to the MFCC-QCP features (FS-4). From the combination of feature sets (FS-8 to FS-12), it can be clearly seen that there exists an improvement in performance for all the combinations. This indicates the complementary information among the feature sets. Further, it is observed that the combination of MFCC-ZFF and MFCC-QCP (FS-9), and the combination of all glottal source features (FS-10) gave the highest detection performance. Also, it can be observed that the combination of conventional MFCC and PLP (FS-11) features showed an improvement in performance, which indicates the presence of complementary information between these features. Overall, the best performance was observed when all glottal source feature sets (FS-1 to FS-5) and conventional MFCC and PLP features (FS-6 and FS-7) were combined. The combination of all the feature sets (FS-12) gave an accuracy of 78.37\% , AUC of 0.84, and EER of 0.207, which highlights the complementary nature of the conventional features with the glottal source features for voice pathology detection.

\begin{table}[h]
\centering
\caption{\label{HUPA_a} Results for the vowel /a/ in the HUPA database with individual feature sets (FS-1 to FS-7) and combination of feature sets (FS-8 to FS-12). SE refers to sensitivity, SP refers to specificity, AUC refers to area under the ROC curve and EER refers to equal error rate.}
\vspace{1mm}
\label{accuracies} 
\small\addtolength{\tabcolsep}{-5pt}
\resizebox{7.5cm}{3.2cm}{
\begin{tabular}{|l|c|c|c  |c|c|} 
\hline
Feature set & {{Accuracy}} [\%]& {{SE}} & {{SP}} & {{AUC}} & {{EER}}\\ \hline  \hline 
FS-1   	& 68.51$\pm$4.80    & 0.76   & 0.62    & 0.72 & 0.304\\
FS-2	& 66.89$\pm$3.54    & 0.71   & 0.63    & 0.71 & 0.315\\
FS-3	& 64.33$\pm$4.92    & 0.78   & 0.55    & 0.71 & 0.322\\
FS-4	& 70.28$\pm$4.14    & 0.75   & 0.66    & 0.77 & 0.267\\
FS-5	& {\bf72.89}$\pm$4.02    & 0.85   & 0.63    & {\bf0.78} & {\bf0.253}\\ 
FS-6	& 72.00$\pm$4.36    & 0.80   & 0.63    & 0.73 & 0.279\\ 
FS-7	& 72.36$\pm$4.23    & 0.78   & 0.67    & 0.74 & 0.272\\ \hline \hline
FS-8	& 73.74$\pm$4.12    & 0.74   & 0.73    & 0.78 & 0.249\\ 
FS-9	& 76.67$\pm$5.01    & 0.81   & 0.72    & {\bf0.79} & 0.241\\ 
FS-10	& {\bf76.96}$\pm$4.12    & 0.80   & 0.74    & {\bf0.79} & {\bf0.238}\\ 
FS-11	& 73.68$\pm$4.22    & 0.78   & 0.69    & 0.76 & 0.249\\ \hline \hline
FS-12	& {\bf78.37}$\pm$4.18    & 0.80   & 0.77    & {\bf0.84} & {\bf0.207}\\ \hline 
\end{tabular}}
\end{table}

Tables~\ref{SVD_a},~\ref{SVD_i}~and~\ref{SVD_u} show the detection results for the vowels /a/, /i/ and /u/ in the SVD database with the individual feature sets (FS-1 to FS-7) and combination of feature sets (FS-8 to FS-12). 
%NOTE: THE FOLLOWING SENTENCE IS UNCLEAR: As in the results obtained for the HUPA database, also here the individual feature sets (FS-1 to FS-7), MFCC-ZFF features (FS-5) provides better performance in terms of AUC for all the cases (vowels /a/, /i/ and /u/). 
As in the results obtained for the HUPA database, here also (in the case of individual feature sets FS-1 to FS-7), the MFCC-ZFF features (FS-5) provided the best performance in terms of AUC and EER in all vowels.
%For the vowel /i/ (Table~\ref{SVD_i}), conventional MFCC features provided the same AUC (0.69) as MFCC-ZFF. 
In terms of accuracy, the PLP features gave the highest performance for all the three vowels. Complementary information among the features can be observed from the results of feature sets from FS-8 to FS-12. The combination of all the glottal source features (FS-10) gave the highest AUC and lowest EER for all the three vowels. Overall, the best performance was achieved when all the glottal source feature sets were combined with the conventional MFCC and PLP features (i.e., FS-12). The combination of all the feature sets gave an AUC of 0.76, 0.75 and 0.73, EER of 0.271, 0.284, and 0.292, and accuracy of 74.32\%, 72.29\% and 71.45\% for the vowels /a/, /i/ and /u/, respectively. It should be noted that the detection experiments with the vowel /a/ achieved better results (highest AUC, accuracy and lowest EER) compared to the vowels /i/ and /u/. This might be due to the use of wider vocal tract openings in the production of the vowel /a/ which results in less interaction between the glottal source and the vocal tract \cite{GarciaMG19}. This observation is also in line with results reported in \cite{GarciaMG19}.

\begin{table}[h]
\centering
\caption{{\label{SVD_a}Results for the vowel /a/ in the SVD database with individual feature sets (FS-1 to FS-7) and combination of feature sets (FS-8 to FS-12). SE refers to sensitivity, SP refers to specificity, AUC refers to area under the ROC curve and EER refers to equal error rate.}}
\vspace{1mm}
\label{accuracies} 
\small\addtolength{\tabcolsep}{-5pt}
\resizebox{7.5cm}{3.2cm}{
\begin{tabular}{|l|c|c|c  |c|c|} 
\hline
Feature set & {{Accuracy}} [\%]& {{SE}} & {{SP}} & {{AUC}} & {{EER}}\\ \hline  \hline 
FS-1   	& 65.91$\pm$4.82    &0.70    &0.62    & 0.68 & 0.341\\
FS-2	& 66.76$\pm$4.19    &0.72    &0.62    & 0.70 & 0.332\\
FS-3	& 64.11$\pm$4.57    &0.67    &0.61    & 0.71 & 0.338\\
FS-4	& 65.12$\pm$4.28    &0.68    &0.63    & 0.72 & 0.329\\
FS-5	& 66.16$\pm$4.92    &0.69    &0.63    & {\bf0.74} & {\bf0.307}\\ 
FS-6	& 67.25$\pm$4.11    &0.70    &0.64    & 0.72 & 0.315\\ 
FS-7	& {\bf68.19}$\pm$3.84    &0.70    &0.65    & 0.73 & 0.320\\ \hline \hline
FS-8	& 68.64$\pm$4.02    &0.69    &0.66    & 0.73 & 0.319\\ 
FS-9	& 68.12$\pm$4.12    &0.70    &0.65    & 0.74 & 0.297\\ 
FS-10	& {\bf72.12}$\pm$4.35    &0.73    &0.69    & {\bf0.75} & {\bf0.286}\\ 
FS-11	& 69.97$\pm$4.02    &0.71    &0.69    & 0.74 & 0.302\\ \hline \hline
FS-12	& {\bf74.32}$\pm$3.58    &0.75    &0.71    & {\bf0.76} & {\bf0.271}\\ \hline 
\end{tabular}}
\end{table}

\begin{table}[h]
\centering
\caption{{\label{SVD_i}Results for the vowel /i/ in the SVD database with individual feature sets (FS-1 to FS-7) and combination of feature sets (FS-8 to FS-12).SE refers to sensitivity, SP refers to specificity, AUC refers to area under the ROC curve and EER refers to equal error rate.}}
\vspace{1mm}
\label{accuracies} 
\small\addtolength{\tabcolsep}{-5pt}
\resizebox{7.5cm}{3.2cm}{
\begin{tabular}{|l|c|c|c  |c|c|} 
\hline
Feature set & {{Accuracy}} [\%]& {{SE}} & {{SP}} & {{AUC}} & {{EER}}\\ \hline  \hline 

FS-1   	& 63.57$\pm$4.91    & 0.68   &0.60    & 0.67  & 0.361\\
FS-2	& 65.12$\pm$4.73    & 0.70   &0.60    & 0.67  & 0.353\\
FS-3	& 63.11$\pm$4.89    & 0.66   &0.59    & 0.66  & 0.367\\
FS-4	& 65.51$\pm$4.48    & 0.69   &0.61    & 0.67  & 0.359\\
FS-5	& 66.10$\pm$4.16    & 0.70   &0.62    & {\bf0.70}  & {\bf0.322}\\ 
FS-6	& 66.81$\pm$4.22    & 0.71   &0.61    & {\bf0.70}  & 0.323\\ 
FS-7	& {\bf67.79}$\pm$4.54    & 0.71   &0.63    & 0.69  & 0.334\\ \hline \hline
FS-8	& 67.04$\pm$4.63    & 0.69   &0.64    & 0.70  & 0.326\\ 
FS-9	& 67.68$\pm$4.32    & 0.70   &0.62    & {\bf0.71}  & 0.316\\ 
FS-10	& 66.91$\pm$4.27    & 0.69   &0.63    & {\bf0.71}  & {\bf0.310}\\ 
FS-11	& {\bf69.11}$\pm$3.97    & 0.72   &0.64    & 0.70  & 0.324\\ \hline \hline
FS-12	& {\bf72.29}$\pm$4.02    & 0.74   &0.67    & {\bf0.75}  & {\bf0.284}\\ \hline 
\end{tabular}}
\end{table}

\begin{table}[h]
\centering
\caption{{\label{SVD_u}Results for the vowel /u/ in the SVD database with individual feature sets (FS-1 to FS-7) and combination of feature sets (FS-8 to FS-12).SE refers to sensitivity, SP refers to specificity, AUC refers to area under the ROC curve and EER refers to equal error rate.}}
\vspace{1mm}
\label{accuracies} 
\small\addtolength{\tabcolsep}{-5pt}
\resizebox{7.5cm}{3.2cm}{
\begin{tabular}{|l|c|c|c  |c|c|} 
\hline
Feature set & {{Accuracy}} [\%]& {{SE}} & {{SP}} & {{AUC}} & {{EER}}\\ \hline  \hline 
FS-1   	& 63.12$\pm$4.22    &0.65    &0.61    &0.66   & 0.361\\
FS-2	& 63.59$\pm$4.35    &0.64    &0.63    &0.66   & 0.365\\
FS-3	& 62.11$\pm$4.08    &0.62    &0.61    &0.64   & 0.378\\
FS-4	& 65.91$\pm$3.94    &0.69    &0.62    &0.67   & 0.345\\
FS-5	& 65.77$\pm$4.19    &0.70    &0.61    &{\bf0.68}   & {\bf0.336}\\ 
FS-6	& 65.49$\pm$4.32    &0.69    &0.60    &0.67   & 0.349\\ 
FS-7	& {\bf67.14}$\pm$3.92    &0.71    &0.62    &{\bf0.68}   & 0.343\\ \hline \hline
FS-8	& 64.98$\pm$3.96    &0.69    &0.61    &0.68   & 0.347\\ 
FS-9	& 66.89$\pm$3.48    &0.70    &0.63    &0.68   & 0.330\\ 
FS-10	& {\bf68.84}$\pm$3.67    &0.71    &0.65    &{\bf0.71}   & {\bf0.319}\\ 
FS-11	& 68.22$\pm$3.84    &0.70    &0.66    &0.69   & 0.328\\ \hline \hline
FS-12	& {\bf71.45}$\pm$3.79    &0.72    &0.69    &{\bf0.73}   & {\bf0.292}\\ \hline 
\end{tabular}}
\end{table}

\begin{table}[h]
\centering
\caption{{\label{SVD_rs}Results for the sentences in the SVD database with individual feature sets (FS-1 to FS-7) and combination of feature sets (FS-8 to FS-12).SE refers to sensitivity, SP refers to specificity, AUC refers to area under the ROC curve and EER refers to equal error rate.}}
\vspace{1mm}
\label{accuracies} 
\small\addtolength{\tabcolsep}{-5pt}
\resizebox{7.5cm}{3.2cm}{
\begin{tabular}{|l|c|c|c  |c|c|} 
\hline
Feature set & {{Accuracy}} [\%]& {{SE}} & {{SP}} & {{AUC}} & {{EER}}\\ \hline  \hline 
FS-1   	& 68.14$\pm$4.18    &0.68    &0.67    &0.67   & 0.332\\
FS-2	& 65.28$\pm$4.35    &0.69    &0.61    &0.65   & 0.343\\
FS-3	& 64.47$\pm$4.05    &0.62    &0.66    &0.65   & 0.345\\
FS-4	& 65.95$\pm$3.92    &0.64    &0.67    &0.68   & 0.327\\
FS-5	& {\bf68.92}$\pm$3.84    &0.66    &0.71    &{\bf0.70}   & {\bf0.318}\\ 
FS-6	& 67.89$\pm$4.20    &0.65    &0.69    &0.68   & 0.331\\ 
FS-7	& 67.52$\pm$4.58    &0.65    &0.68    &0.70   & 0.322\\ \hline \hline
FS-8	& 70.81$\pm$3.81    &0.71    &0.70    &{\bf0.74}   & {\bf0.286}\\ 
FS-9	& 72.15$\pm$3.98    &0.69    &0.73    &0.73   & 0.307\\ 
FS-10	& {\bf73.72}$\pm$4.03    &0.69    &0.75    &0.73   & 0.296\\ 
FS-11	& 69.93$\pm$3.79    &0.68    &0.70    &0.72   & 0.309\\ \hline \hline
FS-12	& {\bf76.19}$\pm$3.62    &0.72    & 0.78   &{\bf0.78}   & {\bf0.262}\\ \hline 
\end{tabular}}
\end{table}

Table~\ref{SVD_rs} shows the results for the continuous speech sentences in the SVD database with individual feature sets (FS-1 to FS-7) and combination of feature sets (FS-8 to FS-12). In the case of individual feature sets, MFCC-ZFF (FS-5) achieved the highest AUC of 0.7 and lowest EER of 0.318. Conventional MFCCs (FS-6), the proposed MFCC-QCP (FS-4) and MFCC-ZFF (FS-5) had nearly similar performance. %In terms of accuracy, MFCC-ZFF features provided higher accuracy than any other individual feature set. 
The results of the combination of feature sets (FS-8 to FS-12) indicate the complementary nature of the feature sets. In the case of combination of feature sets, 1-dimensional glottal source features (combination of QCP features, ZFF features and features derived directly from voice signals) gave the highest AUC of 0.74 and lowest EER of 0.286. Overall, the best performance was achieved (EER of 0.262, AUC of 0.78 and accuracy of 76.19\%) when all the feature sets were combined, indicating the complementary nature of information of the glottal source features with the existing conventional spectral features, MFCCs and PLPs. 

It is worth noting that there exist studies in the literature \cite{GOMEZGARCIA2019,HEGDE2018,MSMFCC} which report detection performance superior to that obtained in this study, but many of those studies have only included a small portion of the database and/or limited the analyses to a restricted number of pathologies. It is observed that the trend in the results reported in this paper are in line with the results reported in \cite{GarciaMG19,svduse2}.

\section{Summary}
\label{sec:summary}
Glottal source features were studied in this work in the analysis of normal and pathological voice, and these features were further used in voice pathology detection. The glottal source features were derived from three signals: from the glottal flows estimated with the QCP inverse filtering method, from the approximate source signals computed with the ZFF method and directly from acoustic voice signals. Analysis of features revealed that glottal source features help in discriminating normal voice from pathological voice. %contains the significant discrimination capability. Also two new feature sets MFCC-QCP and MFCC-ZFF were proposed which are derived from QCP and ZFF methods. 
Detection experiments were carried out using two databases with individual glottal source feature sets and with a combination of features. Experiments showed that on their own the studied glottal source features provide better discrimination compared to spectral features such as MFCCs and PLPs features. Also, it was shown that complementary information exists among the different glottal source features. 
%In several cases, performance with the proposed glottal source features is better than the existing spectral features such as MFCC and PLP features. 
Further, the combination of the existing spectral features with the glottal source features resulted in improved detection performance, indicating the complementary nature of features. 

Motivated by the voice pathology detection performance achieved using glottal source features, we intend to use these features in the future for the classification of pathologies and for predicting the level of pathology (i.e., quantifying the severity level, for example, as mild, medium, high and very high), which may be helpful for diagnosis.  

% \bibliographystyle{unsrt}
% \bibliography{Pathology.bib,Phonations.bib,GA_SFF.bib,bibliography.bib,ksrm_myReferences.bib,refs.bib,ERS.bib,GA_SFF_ZTW.bib,glottalFLOW_ref.bib,Revision_oct17.bib}

\begin{thebibliography}{100}

\bibitem{Disordersbook}
AE~Aronson.
\newblock {\em Clinical Voice Disorders; An Interdisciplinary Approach}.
\newblock Thieme Inc, 1985.

\bibitem{williams2003}
Nelson~R Williams.
\newblock Occupational groups at risk of voice disorders: a review of the literature.
\newblock {\em Occupational Medicine}, 53(7):456--460, 2003.

\bibitem{AriasITBE11}
Juli{\'{a}}n~D. Arias{-}Londo{\~{n}}o, Juan~Ignacio Godino{-}Llorente, Nicol{\'{a}}s S{\'{a}}enz{-}Lech{\'{o}}n, V{\'{\i}}ctor Osma{-}Ruiz, and Germ{\'{a}}n Castellanos{-}Dom{\'{\i}}nguez.
\newblock Automatic detection of pathological voices using complexity measures, noise parameters, and mel-cepstral coefficients.
\newblock {\em {IEEE} Transactions on Biomedical Engineering}, 58(2):370--379, 2011.

\bibitem{GOMEZGARCIA2019}
Jorge Andr{\'{e}}s~G{\'{o}}mez Garc{\'{\i}}a, Laureano Moro{-}Vel{\'{a}}zquez, and Juan~Ignacio Godino{-}Llorente.
\newblock On the design of automatic voice condition analysis systems. part {I:} review of concepts and an insight to the state of the art.
\newblock {\em Biomedical Signal Processing and Control}, 51:181 -- 199, 2019.

\bibitem{little2009suitability}
Max~A Little, Patrick~E McSharry, Eric~J Hunter, Jennifer Spielman, and Lorraine~O Ramig.
\newblock Suitability of dysphonia measurements for telemonitoring of parkinson's disease.
\newblock {\em IEEE Transactions on Biomedical Engineering}, 56(4):1015, 2009.

\bibitem{Silva2009}
D\'{a}rcio~G. Silva, Lu\'{\i}s~C. Oliveira, and M\'{a}rio Andrea.
\newblock Jitter estimation algorithms for detection of pathological voices.
\newblock {\em EURASIP Journal on Advances in Signal Processing}, pages 9:1--9:9, Jan. 2009.

\bibitem{jittersty}
M~Vasilakis and Y~Stylianou.
\newblock Voice pathology detection based on short-term jitter estimations in running speech.
\newblock {\em Folia Phoniatrica Logopedica}, 61(3):153--170, 2009.

\bibitem{PerturbationNDA}
Yu~Zhang, Jack~J. Jiang, Laura Biazzo, and Malinda Jorgensen.
\newblock Perturbation and nonlinear dynamic analyses of voices from patients with unilateral laryngeal paralysis.
\newblock {\em Journal of Voice}, 19(4):519 -- 528, 2005.

\bibitem{pertparsa}
Vijay Parsa and Donald~G. Jamieson.
\newblock Acoustic discrimination of pathological voice.
\newblock {\em Journal of Speech, Language, and Hearing Research}, 44(2):327--339, 2001.

\bibitem{NFL}
J.~R. {Orozco-Arroyave}, E.~A. {Belalcazar-Bolaños}, J.~D. {Arias-Londoño}, J.~F. {Vargas-Bonilla}, S.~{Skodda}, J.~{Rusz}, K.~{Daqrouq}, F.~{Hönig}, and E.~{Nöth}.
\newblock Characterization methods for the detection of multiple voice disorders: Neurological, functional, and laryngeal diseases.
\newblock {\em IEEE Journal of Biomedical and Health Informatics}, 19(6):1820--1828, Nov 2015.

\bibitem{MEKYSKArca}
Jiri Mekyska, Eva Janousova, Pedro Gomez-Vilda, Zdenek Smekal, Irena Rektorova, Ilona Eliasova, Milena Kostalova, Martina Mrackova, Jesus~B. Alonso-Hernandez, Marcos Faundez-Zanuy, and Karmele~López de~Ipiña.
\newblock Robust and complex approach of pathological speech signal analysis.
\newblock {\em Neurocomputing}, 167:94 -- 111, 2015.

\bibitem{roy2013evidence}
Nelson Roy, Julie Barkmeier-Kraemer, Tanya Eadie, M~Preeti Sivasankar, Daryush Mehta, Diane Paul, and Robert Hillman.
\newblock Evidence-based clinical voice assessment: A systematic review.
\newblock {\em American Journal of Speech-Language Pathology}, 2013.

\bibitem{GarciaMG19}
Jorge Andr{\'{e}}s~G{\'{o}}mez Garc{\'{\i}}a, Laureano Moro{-}Vel{\'{a}}zquez, and Juan~Ignacio Godino{-}Llorente.
\newblock On the design of automatic voice condition analysis systems. part {II:} review of speaker recognition techniques and study on the effects of different variability factors.
\newblock {\em Biomedical Signal Processing and Control}, 48:128--143, 2019.

\bibitem{MANFREDIf0}
C.~Manfredi, M.~D'Aniello, P.~Bruscaglioni, and A.~Ismaelli.
\newblock A comparative analysis of fundamental frequency estimation methods with application to pathological voices.
\newblock {\em Medical Engineering \& Physics}, 22(2):135 -- 147, 2000.

\bibitem{speccep}
Christopher~R. Watts and Shaheen~N. Awan.
\newblock Use of spectral/cepstral analyses for differentiating normal from hypofunctional voices in sustained vowel and continuous speech contexts.
\newblock {\em Journal of Speech, Language, and Hearing Research}, 54(6):1525--1537, 2011.

\bibitem{patel2018recommended}
Rita~R Patel, Shaheen~N Awan, Julie Barkmeier-Kraemer, Mark Courey, Dimitar Deliyski, Tanya Eadie, Diane Paul, Jan~G {\v{S}}vec, and Robert Hillman.
\newblock Recommended protocols for instrumental assessment of voice: American speech-language-hearing association expert panel to develop a protocol for instrumental assessment of vocal function.
\newblock {\em American Journal of Speech-Language Pathology}, 27(3):887--905, 2018.

\bibitem{hnrjasa}
Yingyong Qi and Robert~E Hillman.
\newblock Temporal and spectral estimations of harmonics-to-noise ratio in human voice signals.
\newblock {\em The Journal of the Acoustical Society of America}, 102(1):537--543, 1997.

\bibitem{hnricassp}
J.~{Lee}, S.~{Kim}, and H.~{Kang}.
\newblock Detecting pathological speech using contour modeling of harmonic-to-noise ratio.
\newblock In {\em ICASSP}, pages 5969--5973, May 2014.

\bibitem{kasuya1986normalized}
Hideki Kasuya, Shigeki Ogawa, Kazuhiko Mashima, and Satoshi Ebihara.
\newblock Normalized noise energy as an acoustic measure to evaluate pathologic voice.
\newblock {\em The Journal of the Acoustical Society of America}, 80(5):1329--1334, 1986.

\bibitem{gne2010}
Juan~Ignacio Godino-Llorente, V{\'\i}ctor Osma-Ruiz, Nicol{\'a}s S{\'a}enz-Lech{\'o}n, Pedro G{\'o}mez-Vilda, Manuel Blanco-Velasco, and Fernando Cruz-Rold{\'a}n.
\newblock The effectiveness of the glottal to noise excitation ratio for the screening of voice disorders.
\newblock {\em Journal of Voice}, 24(1):47--56, 2010.

\bibitem{gneparsa}
Vijay Parsa and Donald~G. Jamieson.
\newblock Identification of pathological voices using glottal noise measures.
\newblock {\em Journal of Speech, Language, and Hearing Research}, 43(2):469--485, 2000.

\bibitem{gneacta}
D.~Michaelis, T.~Gramss, and H.~W. Strube.
\newblock Glottal-to-noise excitation ratio a new measure for describing pathological voices.
\newblock {\em Acta Acustica united with Acustica}, 83(4):700--706, 1997.

\bibitem{specmaveba}
Rub{\'{e}}n Fraile, Juan~Ignacio Godino{-}Llorente, Nicol{\'{a}}s S{\'{a}}enz{-}Lech{\'{o}}n, Juana~M. Guti{\'{e}}rrez{-}Arriola, and V{\'{\i}}ctor Osma{-}Ruiz.
\newblock Spectral analysis of pathological voices: sustained vowels vs running speech.
\newblock In {\em MAVEBA}, pages 67--70, 2011.

\bibitem{cepfeat}
A.~{Benba}, A.~{Jilbab}, and A.~{Hammouch}.
\newblock Discriminating between patients with parkinson’s and neurological diseases using cepstral analysis.
\newblock {\em IEEE Transactions on Neural Systems and Rehabilitation Engineering}, 24(10):1100--1108, Oct 2016.

\bibitem{mfccann}
Juan~Ignacio Godino{-}Llorente and Pedro~G{\'{o}}mez Vilda.
\newblock Automatic detection of voice impairments by means of short-term cepstral parameters and neural network based detectors.
\newblock {\em {IEEE} Transactions on Biomedical Engineering}, 51(2):380--384, 2004.

\bibitem{LPfeatpat}
Juan~Ignacio Godino{-}Llorente, Santiago Aguilera{-}Navarro, and Pedro~G{\'{o}}mez Vilda.
\newblock Lpc, {LPCC} and {MFCC} parameterisation applied to the detection of voice impairments.
\newblock In {\em INTERSPEECH}, pages 965--968, 2000.

\bibitem{featclassifier}
Jennifer~C. Saldanha, T.~Ananthakrishna, and Rohan Pinto.
\newblock Vocal fold pathology assessment using mel-frequency cepstral coefficients and linear predictive cepstral coefficients features.
\newblock {\em Journal of Medical Imaging and Health Informatics}, 4(2):168--173, 2014.

\bibitem{LITTLEnda}
Max~A. Little, Declan~A.E. Costello, and Meredydd~L. Harries.
\newblock Objective dysphonia quantification in vocal fold paralysis: Comparing nonlinear with classical measures.
\newblock {\em Journal of Voice}, 25(1):21 -- 31, 2011.

\bibitem{plpextract}
Hynek Hermansky.
\newblock Perceptual linear predictive (plp) analysis of speech.
\newblock {\em The Journal of the Acoustical Society of America}, 87(4):1738--1752, 1990.

\bibitem{hillenbrand1994acoustic}
James Hillenbrand, Ronald~A Cleveland, and Robert~L Erickson.
\newblock Acoustic correlates of breathy vocal quality.
\newblock {\em Journal of Speech, Language, and Hearing Research}, 37(4):769--778, 1994.

\bibitem{CPPbio}
Rub{\'{e}}n Fraile, Juan~Ignacio Godino{-}Llorente, Nicol{\'{a}}s S{\'{a}}enz{-}Lech{\'{o}}n, V{\'{\i}}ctor Osma{-}Ruiz, and Pedro~G{\'{o}}mez Vilda.
\newblock Use of cepstrum-based parameters for automatic pathology detection on speech - analysis of performance and theoretical justification.
\newblock In {\em {BIOSIGNALS}}, pages 85--91, 2008.

\bibitem{FraileG14}
Rub{\'{e}}n Fraile and Juan~Ignacio Godino{-}Llorente.
\newblock Cepstral peak prominence: {A} comprehensive analysis.
\newblock {\em Biomedical Signal Processing and Control}, 14:42--54, 2014.

\bibitem{drugman2009mutual}
Thomas Drugman, Thomas Dubuisson, and Thierry Dutoit.
\newblock On the mutual information between source and filter contributions for voice pathology detection.
\newblock In {\em INTERSPEECH}, 2009.

\bibitem{FRAILE201311}
Rubén Fraile, Juan~Ignacio Godino-Llorente, Nicolás Sáenz-Lechón, Víctor Osma-Ruiz, and Juana~María Gutiérrez-Arriola.
\newblock Characterization of dysphonic voices by means of a filterbank-based spectral analysis: Sustained vowels and running speech.
\newblock {\em Journal of Voice}, 27(1):11 -- 23, 2013.

\bibitem{tfauma}
K.~{Umapathy}, S.~{Krishnan}, V.~{Parsa}, and D.~G. {Jamieson}.
\newblock Discrimination of pathological voices using a time-frequency approach.
\newblock {\em IEEE Transactions on Biomedical Engineering}, 52(3):421--430, March 2005.

\bibitem{Ghoraani2009}
Behnaz Ghoraani and Sridhar Krishnan.
\newblock A joint time-frequency and matrix decomposition feature extraction methodology for pathological voice classification.
\newblock {\em EURASIP Journal on Advances in Signal Processing}, (1), Sep 2009.

\bibitem{wavelet1}
Roozbeh Behroozmand and Farshad Almasganj.
\newblock Optimal selection of wavelet-packet-based features using genetic algorithm in pathological assessment of patients’ speech signal with unilateral vocal fold paralysis.
\newblock {\em Computers in Biology and Medicine}, 37(4):474 -- 485, 2007.

\bibitem{waveletann}
C.~D.~P. {Crovato} and A.~{Schuck}.
\newblock The use of wavelet packet transform and artificial neural networks in analysis and classification of dysphonic voices.
\newblock {\em IEEE Transactions on Biomedical Engineering}, 54(10):1898--1900, Oct 2007.

\bibitem{fonseca2007wavelet}
Everthon~Silva Fonseca, Rodrigo~Capobianco Guido, Paulo~Rog{\'e}rio Scalassara, Carlos~Dias Maciel, and Jos{\'e}~Carlos Pereira.
\newblock Wavelet time-frequency analysis and least squares support vector machines for the identification of voice disorders.
\newblock {\em Computers in Biology and Medicine}, 37(4):571--578, 2007.

\bibitem{waveletpacket}
Ali Akbari and Meisam~Khalil Arjmandi.
\newblock An efficient voice pathology classification scheme based on applying multi-layer linear discriminant analysis to wavelet packet-based features.
\newblock {\em Biomedical Signal Processing and Control}, 10:209 -- 223, 2014.

\bibitem{MSyan}
M.~{Markaki} and Y.~{Stylianou}.
\newblock Using modulation spectra for voice pathology detection and classification.
\newblock In {\em EMBC}, pages 2514--2517, Sep. 2009.

\bibitem{MarkakiSAG10}
Maria~E. Markaki, Yannis Stylianou, Juli{\'{a}}n~D. Arias{-}Londo{\~{n}}o, and Juan~Ignacio Godino{-}Llorente.
\newblock Dysphonia detection based on modulation spectral features and cepstral coefficients.
\newblock In {\em ICASSP}, pages 5162--5165, 2010.

\bibitem{modspebio}
Laureano Moro-Velázquez, Jorge~Andrés Gómez-García, and Juan~Ignacio Godino-Llorente.
\newblock Voice pathology detection using modulation spectrum-optimized metrics.
\newblock {\em Frontiers in Bioengineering and Biotechnology}, 4:1, 2016.

\bibitem{Kaleem2013}
Muhammad Kaleem, Behnaz Ghoraani, Aziz Guergachi, and Sridhar Krishnan.
\newblock Pathological speech signal analysis and classification using empirical mode decomposition.
\newblock {\em Medical {\&} Biological Engineering {\&} Computing}, 51(7):811--821, Jul 2013.

\bibitem{little2007exploiting}
Max~A Little, Patrick~E McSharry, Stephen~J Roberts, Declan~AE Costello, and Irene~M Moroz.
\newblock Exploiting nonlinear recurrence and fractal scaling properties for voice disorder detection.
\newblock {\em Biomedical engineering online}, 6(1):23, 2007.

\bibitem{ndareview}
Juan Rafael Orozco~Arroyave, Jesus Francisco Vargas~Bonilla, and Edilson Delgado~Trejos.
\newblock Acoustic analysis and non linear dynamics applied to voice pathology detection: A review.
\newblock {\em Recent Patents on Signal Processing}, 2(2):96--107, 2012.

\bibitem{henriquez2009characterization}
Patricia Henr{\'{\i}}quez, Jes{\'{u}}s~B. Alonso, Miguel~A. Ferrer, Carlos~M. Travieso, Juan~Ignacio Godino{-}Llorente, and Fernando D{\'{\i}}az{-}de{-}Mar{\'{\i}}a.
\newblock Characterization of healthy and pathological voice through measures based on nonlinear dynamics.
\newblock {\em IEEE Transactions on Audio, Speech, and Language Processing}, 17(6):1186--1195, 2009.

\bibitem{pean2000fractal}
V~P{\'e}an, M~Ouayoun, C~Fugain, B~Meyer, and CH~Chouard.
\newblock A fractal approach to normal and pathological voices.
\newblock {\em Acta Otolaryngologica}, 120(2):222--224, 2000.

\bibitem{tsanaspd1}
Athanasios Tsanas, Max~A. Little, Patrick~E. McSharry, and Lorraine~O. Ramig.
\newblock Nonlinear speech analysis algorithms mapped to a standard metric achieve clinically useful quantification of average parkinson's disease symptom severity.
\newblock {\em Journal of The Royal Society Interface}, 8(59):842--855, 2011.

\bibitem{VAZInda}
Ghazaleh Vaziri, Farshad Almasganj, and Roozbeh Behroozmand.
\newblock Pathological assessment of patients’ speech signals using nonlinear dynamical analysis.
\newblock {\em Computers in Biology and Medicine}, 40(1):54 -- 63, 2010.

\bibitem{TRAVIESnda}
Carlos~M. Travieso, Jesús~B. Alonso, J.R. Orozco-Arroyave, J.F. Vargas-Bonilla, E.~Nöth, and Antonio~G. Ravelo-García.
\newblock Detection of different voice diseases based on the nonlinear characterization of speech signals.
\newblock {\em Expert Systems with Applications}, 82:184 -- 195, 2017.

\bibitem{giovanni1999determination}
Antoine Giovanni, Maurice Ouaknine, and Jean-Michel Triglia.
\newblock Determination of largest lyapunov exponents of vocal signal: application to unilateral laryngeal paralysis.
\newblock {\em Journal of Voice}, 13(3):341--354, 1999.

\bibitem{tsanaspd}
A.~{Tsanas}, M.~A. {Little}, P.~E. {McSharry}, and L.~O. {Ramig}.
\newblock Accurate telemonitoring of parkinson's disease progression by noninvasive speech tests.
\newblock {\em IEEE Transactions on Biomedical Engineering}, 57(4):884--893, April 2010.

\bibitem{hmment}
Julián~D. Arias-Londoño and Juan~I. Godino-Llorente.
\newblock Entropies from markov models as complexity measures of embedded attractors.
\newblock {\em Entropy}, 17(6):3595--3620, 2015.

\bibitem{GarciaGC12}
Jorge Andr{\'{e}}s~G{\'{o}}mez Garc{\'{\i}}a, Juan~Ignacio Godino{-}Llorente, and Germ{\'{a}}n Castellanos{-}Dom{\'{\i}}nguez.
\newblock Influence of delay time on regularity estimation for voice pathology detection.
\newblock In {\em EMBC}, pages 4217--4220, 2012.

\bibitem{Jittermeasure}
Meike Brockmann, Michael~J. Drinnan, Claudio Storck, and Paul~N. Carding.
\newblock Reliable jitter and shimmer measurements in voice clinics: The relevance of vowel, gender, vocal intensity, and fundamental frequency effects in a typical clinical task.
\newblock {\em Journal of Voice}, 25(1):44 -- 53, 2011.

\bibitem{upm49565}
Jorge Andr{\'e}s~G{\'o}mez Garc{\'i}a.
\newblock Contributions to the design of automatic voice quality analysis systems using speech technologies.
\newblock January 2018.

\bibitem{mfccgmm}
Juan~Ignacio Godino{-}Llorente, Pedro~G{\'{o}}mez Vilda, and Manuel Blanco{-}Velasco.
\newblock Dimensionality reduction of a pathological voice quality assessment system based on gaussian mixture models and short-term cepstral parameters.
\newblock {\em {IEEE} Transactions on Biomedical Engineering}, 53(10):1943--1953, 2006.

\bibitem{nn}
S.~{Hadjitodorov}, B.~{Boyanov}, and B.~{Teston}.
\newblock Laryngeal pathology detection by means of class-specific neural maps.
\newblock {\em IEEE Transactions on Information Technology in Biomedicine}, 4(1):68--73, March 2000.

\bibitem{ARJMANDIsvm}
Meisam~Khalil Arjmandi, Mohammad Pooyan, Mohammad Mikaili, Mansour Vali, and Alireza Moqarehzadeh.
\newblock Identification of voice disorders using long-time features and support vector machine with different feature reduction methods.
\newblock {\em Journal of Voice}, 25(6):e275 -- e289, 2011.

\bibitem{chen2007svm}
Wenxi Chen, Ce~Peng, Xin Zhu, Baikun Wan, and Daming Wei.
\newblock Svm-based identification of pathological voices.
\newblock In {\em International Conference of the Engineering in Medicine and Biology Society}, pages 3786--3789, 2007.

\bibitem{hmmpat}
Juli{\'{a}}n~D. Arias{-}Londo{\~{n}}o, Juan~Ignacio Godino{-}Llorente, Nicol{\'{a}}s S{\'{a}}enz{-}Lech{\'{o}}n, V{\'{\i}}ctor Osma{-}Ruiz, and Germ{\'{a}}n Castellanos{-}Dom{\'{\i}}nguez.
\newblock An improved method for voice pathology detection by means of a hmm-based feature space transformation.
\newblock {\em Pattern Recognition}, 43(9):3100--3112, 2010.

\bibitem{alhussein2018voice}
Musaed Alhussein and Ghulam Muhammad.
\newblock Voice pathology detection using deep learning on mobile healthcare framework.
\newblock {\em IEEE Access}, 6:41034--41041, 2018.

\bibitem{alhussein2019automatic}
Musaed Alhussein and Ghulam Muhammad.
\newblock Automatic voice pathology monitoring using parallel deep models for smart healthcare.
\newblock {\em IEEE Access}, 7:46474--46479, 2019.

\bibitem{muhammad2017smart}
Ghulam Muhammad, SK~Md~Mizanur Rahman, Abdulhameed Alelaiwi, and Atif Alamri.
\newblock Smart health solution integrating iot and cloud: A case study of voice pathology monitoring.
\newblock {\em IEEE Communications Magazine}, 55(1):69--73, 2017.

\bibitem{hossain2016healthcare}
M~Shamim Hossain and Ghulam Muhammad.
\newblock Healthcare big data voice pathology assessment framework.
\newblock {\em iEEE Access}, 4:7806--7815, 2016.

\bibitem{HEGDE2018}
Sarika Hegde, Surendra Shetty, Smitha Rai, and Thejaswi Dodderi.
\newblock A survey on machine learning approaches for automatic detection of voice disorders.
\newblock {\em Journal of Voice}, 2018.

\bibitem{glottalpat1}
Leonardo~A. Forero, Manoela Kohler, Marley~M.B.R. Vellasco, and Edson Cataldo.
\newblock Analysis and classification of voice pathologies using glottal signal parameters.
\newblock {\em Journal of Voice}, 30(5):549 -- 556, 2016.

\bibitem{glottalpat2}
Pedro~G{\'{o}}mez Vilda, Roberto Fern{\'{a}}ndez{-}Ba{\'{\i}}llo, Mar{\'{\i}}a Victoria~Rodellar Biarge, Victor~Nieto Lluis, Agust{\'{\i}}n~{\'{A}}lvarez Marquina, Luis~Miguel Mazaira{-}Fern{\'{a}}ndez, Rafael Mart{\'{\i}}nez{-}Olalla, and Juan~Ignacio Godino{-}Llorente.
\newblock Glottal source biometrical signature for voice pathology detection.
\newblock {\em Speech Communication}, 51(9):759--781, 2009.

\bibitem{glottalpat3}
Pedro~G{\'{o}}mez Vilda, Juan~Ignacio Godino{-}Llorente, Agust{\'{\i}}n~{\'{A}}lvarez Marquina, Rafael Mart{\'{\i}}nez{-}Olalla, Victor~Nieto Lluis, and Mar{\'{\i}}a Victoria~Rodellar Biarge.
\newblock Evidence of glottal source spectral features found in vocal fold dynamics.
\newblock In {\em ICASSP}, pages 441--444, 2005.

\bibitem{MUHAMMAD2017156}
Ghulam Muhammad, Mansour Alsulaiman, Zulfiqar Ali, Tamer~A. Mesallam, Mohamed Farahat, Khalid~H. Malki, Ahmed Al-nasheri, and Mohamed~A. Bencherif.
\newblock Voice pathology detection using interlaced derivative pattern on glottal source excitation.
\newblock {\em Biomedical Signal Processing and Control}, 31:156 -- 164, 2017.

\bibitem{Airaksinen2014}
Manu Airaksinen, Tuomo Raitio, Brad Story, and P.~Alku.
\newblock Quasi closed phase glottal inverse filtering analysis with weighted linear prediction.
\newblock {\em IEEE/ACM Transactions on Audio, Speech, and Language Processing}, 22(3):596--607, Mar. 2014.

\bibitem{Murty32}
K.~Sri~Rama Murty and B.~Yegnanarayana.
\newblock Epoch extraction from speech signals.
\newblock {\em IEEE Transactions on Audio, Speech, and Language Processing}, 16(8):1602--1613, Nov. 2008.

\bibitem{Aparat1}
Matti Airas.
\newblock Tkk aparat: An environment for voice inverse filtering and parameterization.
\newblock {\em Logopedics Phoniatrics Vocology}, 33(1):49--64, 2008.

\bibitem{Paavo11}
Paavo Alku.
\newblock Glottal inverse filtering analysis of human voice production-a review of estimation and parameterization methods of the glottal excitation and their applications.
\newblock {\em Sadhana}, 36(5):623--650, 2011.

\bibitem{Phsing}
Sudarsana~Reddy Kadiri and Bayya Yegnanarayana.
\newblock Analysis and detection of phonation modes in singing voice using excitation source features and single frequency filtering cepstral coefficients {(SFFCC)}.
\newblock In {\em INTERSPEECH}, pages 441--445, 2018.

\bibitem{MDQ}
John Kane and Christer Gobl.
\newblock Wavelet maxima dispersion for breathy to tense voice discrimination.
\newblock {\em {IEEE} Transactions on Audio, Speech and Language Processing}, 21(6):1170--1179, 2013.

\bibitem{PS}
John Kane and Christer Gobl.
\newblock Identifying regions of non-modal phonation using features of the wavelet transform.
\newblock In {\em INTERSPEECH}, pages 177--180, 2011.

\bibitem{DegottexG2011msp}
G.~Degottex, A.~Roebel, and X.~Rodet.
\newblock Phase minimization for glottal model estimation.
\newblock {\em IEEE Transactions on Acoustics, Speech and Language Processing}, 19(5):1080--1090, July 2011.

\bibitem{HuberS2012mspd2ix}
S.~Huber, A.~Roebel, and G.~Degottex.
\newblock Glottal source shape parameter estimation using phase minimization variants.
\newblock In {\em INTERSPEECH}, Portland, USA, September 2012.

\bibitem{ModulationSM}
Laureano Moro-Vel{\'a}zquez, Jorge G{\'o}mez-Garc{\'i}a, Juan~Ignacio Godino-Llorente, and Gustavo Andrade-Miranda.
\newblock Modulation spectra morphological parameters: A new method to assess voice pathologies according to the grbas scale.
\newblock In {\em BioMed research international}, 2015.

\bibitem{MSMFCC}
Julián~David Arias-Londoño, Juan~I. Godino-Llorente, Maria Markaki, and Yannis Stylianou.
\newblock On combining information from modulation spectra and mel-frequency cepstral coefficients for automatic detection of pathological voices.
\newblock {\em Logopedics Phoniatrics Vocology}, 36(2):60--69, 2011.

\bibitem{svddb2}
Manfred P{\"u}tzer and William~J. Barry.
\newblock Instrumental dimensioning of normal and pathological phonation using acoustic measurements.
\newblock {\em Clinical Linguistics \& Phonetics}, 22(6):407--420, 2008.

\bibitem{svddb1}
Manfred P{\"u}tzer and William~J. Barry.
\newblock Saarbr{\"u}cken voice database, institute of phonetics, univ. of saarland, 2010.
\newblock \url{http://www.stimmdatenbank.coli.uni-saarland.de/} (Last viewed April 20, 2019).

\bibitem{svduse2}
David Mart{\'i}nez, Eduardo Lleida, Alfonso Ortega, Antonio Miguel, and Jes{\'u}s Villalba.
\newblock Voice pathology detection on the saarbr{\"u}cken voice database with calibration and fusion of scores using multifocal toolkit.
\newblock In {\em Advances in Speech and Language Technologies for Iberian Languages}, pages 99--109, 2012.

\bibitem{Wong79}
D.~Wong, J.~Markel, and A.~Gray.
\newblock Least squares glottal inverse filtering from the acoustic speech waveform.
\newblock {\em IEEE Transactions on Acoustics, Speech, and Signal Processing}, 27:350--355, 1979.

\bibitem{AME}
Paavo Alku, Jouni Pohjalainen, Martti Vainio, Anne-Maria Laukkanen, and Brad~H. Story.
\newblock Formant frequency estimation of high-pitched vowels using weighted linear prediction.
\newblock {\em The Journal of the Acoustical Society of America}, 134(2):1295--1313, 2013.

\bibitem{myspd}
Sudarsana~Reddy Kadiri and B.~Yegnanarayana.
\newblock Speech polarity detection using strength of impulse-like excitation extracted from speech epochs.
\newblock In {\em ICASSP}, pages 5610--5614, 2017.

\bibitem{arcticDatabase}
{CMU-ARCTIC} speech synthesis databases.

\bibitem{Alku96}
P.~Alku and E.~Vilkman.
\newblock Amplitude domain quotient for characterization of the glottal volume velocity waveform estimated by inverse filtering.
\newblock {\em Speech Communication}, 18:131--138, 1996.

\bibitem{Alku2006amplitude}
Paavo Alku, Matti Airas, Eva Bj{\"o}rkner, and Johan Sundberg.
\newblock An amplitude quotient based method to analyze changes in the shape of the glottal pulse in the regulation of vocal intensity.
\newblock {\em The Journal of the Acoustical Society of America}, 120:1052--1062, 2006.

\bibitem{Alku02}
P.~Alku, T.~Backstrom, and E.~Vilkman.
\newblock Normalized amplitude quotient for parameterization of the glottal flow.
\newblock {\em The Journal of the Acoustical Society of America}, 112:701--710, 2002.

\bibitem{Fant95}
G.~Fant.
\newblock The lf-model revisited. transformations and frequency domain analysis.
\newblock {\em Speech Transmission Laboratory Quarterly Progress and Status Report}, 36:119--156, 1995.

\bibitem{Titze92}
I.~Titze and J.~Sundberg.
\newblock Vocal intensity in speakers and singers.
\newblock {\em The Journal of the Acoustical Society of America}, 91:2936--2946, 1992.

\bibitem{VQ_ASP}
Donald~G Childers and CK~Lee.
\newblock Vocal quality factors: Analysis, synthesis, and perception.
\newblock {\em The Journal of the Acoustical Society of America}, 90(5):2394--2410, 1991.

\bibitem{PSP}
Paavo Alku, Helmer Strik, and Erkki Vilkman.
\newblock Parabolic spectral parameter - {A} new method for quantification of the glottal flow.
\newblock {\em Speech Communication}, 22(1):67--79, 1997.

\bibitem{Emo8}
Sudarsana~Reddy Kadiri, P~Gangamohan, Suryakanth~V. Gangashetty, and B.~Yegnanarayana.
\newblock Analysis of excitation source features of speech for emotion recognition.
\newblock In {\em INTERSPEECH}, pages 1324--1328, 2015.

\bibitem{hillerbrand_breathy}
James Hillenbrand and Robert~A. Houde.
\newblock Acoustic correlates of breathy vocal quality: Dysphonic voices and continuous speech.
\newblock {\em Journal of Speech, Language, and Hearing Research}, 39(2):311--321, 1996.

\bibitem{Kadiri2019}
Sudarsana~Reddy Kadiri and Paavo Alku.
\newblock {Mel-Frequency Cepstral Coefficients of Voice Source Waveforms for Classification of Phonation Types in Speech}.
\newblock In {\em Proc. Interspeech 2019}, pages 2508--2512, 2019.

\bibitem{WPCVox}
Juan~Ignacio Godino-Llorente, V{\'i}ctor Osma-Ruiz, Nicol{\'a}s S{\'a}enz-Lech{\'o}n, Ignacio Cobeta-Marco, Ram{\'o}n Gonz{\'a}lez-Herranz, and Carlos Ram{\'i}rez-Calvo.
\newblock Acoustic analysis of voice using wpcvox: a comparative study with multi dimensional voice program.
\newblock {\em European Archives of Oto-Rhino-Laryngology}, 265(4):465--476, Apr 2008.

\bibitem{ROC}
Tom Fawcett.
\newblock An introduction to roc analysis.
\newblock {\em Pattern Recognition Letters}, 27(8):861 -- 874, 2006.

\end{thebibliography}

\begin{IEEEbiography}[{\includegraphics[width=1.5in,height=1.2in,clip,keepaspectratio]{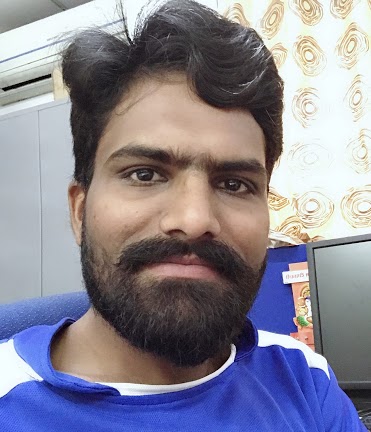}}]{Sudarsana Reddy Kadiri} was born in Kurnool, Andhra Pradesh, India, in 1990. He received his Bachelor of Technology degree from Jawaharlal Nehru Technological University (JNTU), Hyderabad, India, in 2011, with a specialization in Electronics and Communication Engineering (ECE). He did his M.S. (Research) between 2011 and 2014, and later converted to Ph.D. at International Institute of Information Technology, Hyderabad (IIIT-H), India. He obtained his Ph.D. degree from the Department of ECE, IIIT-H, in 2018. He was a teaching assistant for several courses at IIIT-H during 2012-2018. Currently, he is a Postdoctoral researcher with the Department of Signal Processing and Acoustics at Aalto University, Finland. His research interests include signal processing, speech analysis, speech synthesis, paralinguistics, affective computing, voice pathologies, machine learning and auditory neuroscience.
\end{IEEEbiography}

\begin{IEEEbiography}[{\includegraphics[width=1in,height=1.2in]{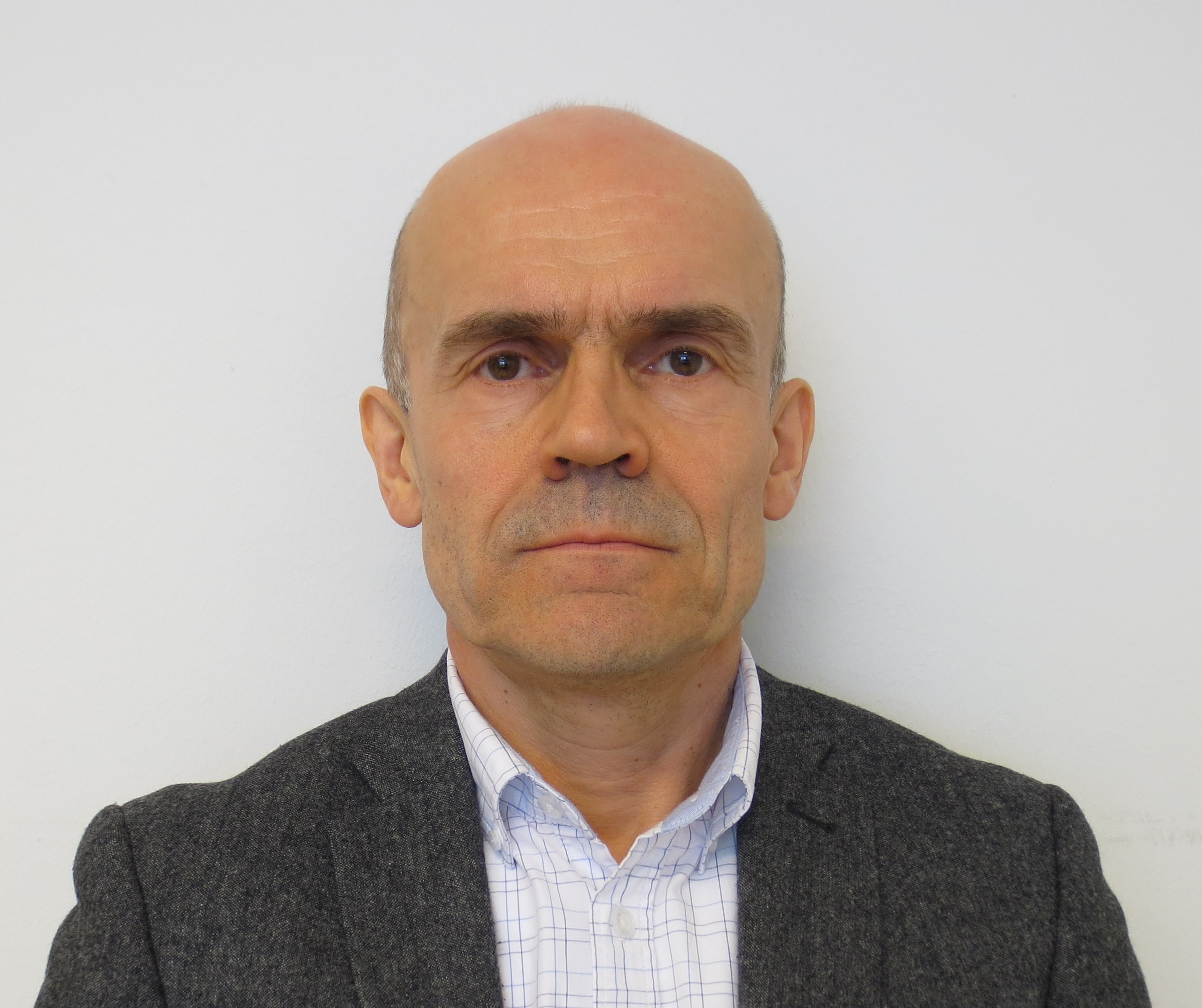}}]{Paavo Alku} received the M.Sc., Lic.Tech., and Dr.Sc.(Tech) degrees from the Helsinki University of Technology, Espoo, Finland, in 1986, 1988, and
1992, respectively. He was an Assistant Professor with the Asian Institute of Technology, Bangkok, Thailand, in 1993, and also with the University of Turku, Finland, from 1994 to 1999. He is currently a Professor of speech communication technology with Aalto University, Espoo. His research interests include analysis and parameterization of speech production, statistical parametric speech synthesis, spectral modeling of speech, speech enhancement, and cerebral processing of speech.

\end{IEEEbiography}

\end{document}